\documentclass[a4paper,10pt]{article}
\usepackage{graphicx}
\usepackage{amssymb}
\usepackage{amsmath}
\usepackage{dcolumn}
\usepackage{bm}
\usepackage{multirow}
\usepackage{cite}
\usepackage{xcolor}
\usepackage{url}
\usepackage{mathrsfs}
\usepackage[normalem]{ulem}
\usepackage{lscape}
\usepackage{soul}
\usepackage{subcaption} 
\usepackage{hyperref}
\usepackage{dcolumn,ulem,enumitem}
\usepackage{bm}
\usepackage{xcolor}
\usepackage{tikz}
\usepackage{booktabs}

\begin{document}

\huge

\begin{center}
Opacity predictions in plasmas under stellar conditions using deep learning
\end{center}

\vspace{0.5cm}

\large

\begin{center}
Djamel Benredjem$^{a,}$\footnote{djamel.benredjem@universite-paris-saclay.fr} and Jean-Christophe Pain$^{b,c}$
\end{center}

\normalsize

\begin{center}
\it $^a$Laboratoire Aim\'e Cotton, Universit\'e Paris-Saclay, Orsay, France\\
\it $^b$CEA, DAM, DIF, F-91297 Arpajon, France\\
\it $^c$Universit\'e Paris-Saclay, CEA, Laboratoire Mati\`ere en Conditions Extr\^emes,\\
\it 91680 Bruy\`eres-le-Ch\^atel, France\\
\end{center}

\vspace{0.5cm}

\begin{abstract}
The aim of this work is to predict the opacity of plasmas under stellar conditions. We focus on iron and nickel, as these elements have been extensively investigated both theoretically and experimentally. In certain regimes, notably under non-local thermodynamic equilibrium, calculating the spectral opacity can be computationally demanding. To mitigate this difficulty, we employ deep learning models to predict accurate opacities while achieving a substantial reduction in computational cost. Specifically, we develop and train a hybrid model that combines a convolutional neural network with a multilayer perceptron. The proposed approach provides both the spectral opacity over the 0--10,000 eV energy range, and the corresponding mean opacities for large sets of temperatures and mass densities, in close agreement with standard computational methods. The agreement is very satisfactory, except in the L- and M-shell energy ranges for nickel at high temperatures, where a very large number of emerging transitions leads to substantial discrepancies. 
\end{abstract}

\section{Introduction}\label{sec:Introduction}
Opacity studies play a central role in understanding both laboratory and stellar plasmas. In this work, we concentrate on iron and nickel. Both elements are widely investigated theoretically as well as experimentally. Iron opacity measurements were conducted at solar interior temperatures using the Z machine at Sandia National Laboratories \cite{Bailey2015,Nagayama2017,Loisel2025}. Nickel is an element of the iron group, and nickel laser-produced plasmas have been investigated at the LULI facility \cite{Dozieres2019}, enabling simultaneous measurements in the X-ray and XUV ranges. In parallel, iron and nickel opacity calculations were carried out under experimental conditions representative of the envelopes of intermediate-mass stars \cite{Gilles2011,Turck2009}.

Iron, nickel, and chromium drive oscillations through the $\kappa$-mechanism in upper main-sequence stars. In Slowly-Pulsating B (SPB) stars, the coolest of these pulsators, this mechanism accounts for high-overtone gravity modes, while in $\beta$ Cephei stars, it excites low-order p- and g-modes. Hybrid pulsators located in the instability strip overlap (near 10~$M_{\odot}$) exhibit both oscillation types. Similar pulsational behavior is also found in B-type subdwarfs beyond the main sequence. Notably, nickel is essential for exciting pulsations, especially high-overtone g-modes in upper main-sequence stars (see for instance Ref. \cite{Hui2022} and references therein).

\indent Opacities issues are also crucial in inertial confinement fusion \cite{Atzeni2004,Abu-Shwareb2024}. In this context, the general non-LTE problem requires calculating the transition rates between levels (or configurations/superconfigurations), and solving a set of collisional-radiative equations. In LTE conditions, one can simply use the Saha-Boltzmann equation to calculate energy level populations and charge state distributions. Hopefully, NLTE radiation-hydrodynamics simulations still often rely, at least in part, on effective temperatures combined with high-resolution LTE opacity tables. Effective temperatures $T_{\rm eff}$ (sometimes called ionization temperatures) \cite{Busquet1993,Bauche2006} are defined so that the LTE spectrum at $T_{\rm eff}$ closely resembles the non-LTE spectrum at $T$. The value of $T_{\rm eff}$ is chosen to ensure that the LTE ionization of the plasma at $T_{\rm eff}$ matches the average non-LTE ionization at $T$.

Accurate radiative properties of carbon, hydrogen, oxygen, and germanium, which enter the composition of the ablator, are therefore essential. The ablator is the outermost layer of the fuel capsule, acting as the primary engine for compression. When high-intensity laser beams (or X-rays in indirect drive) strike the surface, the ablator material heats up rapidly, turns into plasma, and expands outward at high velocity. This outward expansion creates a powerful inward reaction force —an implosion— that crushes the internal deuterium-tritium fuel. To achieve maximum efficiency, the ablator must be engineered with precise density and opacity to ensure the fuel remains cool while being squeezed to the extreme pressures required for nuclear fusion. The ablator is often doped with a few percent of higher-Z materials such as silicon or germanium. The main role of this doping is to modify radiative transport: these elements increase the X-ray opacity of the ablator, thereby limiting fuel preheat by energetic photons (``hard X-rays''). This helps keep the DT fuel colder and denser prior to implosion. The dopant also contributes to stabilizing the implosion hydrodynamics. By tuning energy absorption and deposition within the ablator, it can reduce certain Rayleigh–Taylor instabilities and improve compression symmetry. The choice between silicon, germanium (or other dopants) depends on the hohlraum radiation spectrum and on the desired trade-off between radiative absorption, stability, and ablation efficiency.

As the opacity is a function of the photon energy and depends on temperature and density, investigating laboratory plasmas or stellar atmospheres represents a significant challenge. Traditionally, opacities are computed using complex and computationally expensive physics codes such as, for instance, {\sc atomic} \cite{Fontes2015,Hakel2006}, {\sc sco-rcg} \cite{Porcherot2011}, or {\sc hullac} \cite{BarShalom2001}. Given the large amount of calculation at each time step in each spatial mesh, for the corresponding radiation field, and the temperature-density variations, deep learning offers a particularly fast and efficient alternative for spectral opacity predictions.

Our work focuses on predicting iron and nickel opacity spectra, using deep learning, in the conditions of the above mentioned astrophysical applications. The main objective of our study is not the intrinsic accuracy of opacity models, but their high computational cost. It is worth stressing that detailed opacity calculations require solving complex atomic physics models over fine spectral grids and over large thermodynamic spaces (temperature, density), which makes them expensive when repeated evaluations are needed, for instance in radiation-hydrodynamics simulations. We must emphasize that our goal is to replace these costly evaluations with a surrogate model providing orders-of-magnitude speed-up at inference time, while maintaining good confidence in the results. Although the training phase is computationally demanding, it is performed offline and its cost is amortized over numerous subsequent applications. The training phase requires a large database of precomputed opacity spectra, which is indeed computationally expensive. However, this cost is incurred only once and offline. Once trained, the model provides extremely fast evaluations of opacity. 

To this end, we adopt a hybrid architecture combining a multilayer perceptron (MLP) \cite{Przybyla2024,Abdurrakhman2025} and a convolutional neural network (CNN) \cite{Krichen2023,Back2019,Churchill2020}. The MLP models temperature and density variations, while the CNN captures the dependence on photon energy. A multilayer perceptron, which can be seen as a ``standard'' or fully connected neural network, is able to approximate global dependencies but struggles to reproduce the fine and highly-structured spectral features present in opacity data. In particular, opacity spectra exhibit sharp lines and localized features across energy space that are difficult to capture using only global dense mappings. The CNN is particularly well suited to opacity spectra because it operates through local convolutional kernels that act as learned feature detectors in energy space. These kernels efficiently capture sharp spectral lines, line broadening, and local correlations between nearby energies. This allows the model to naturally reproduce the fine spectral structure that is difficult to represent with fully connected architectures alone. Recently, CNN learning has been applied to spectral opacities in NLTE laser plasmas \cite{Kluth2020} and to the emission of Balmer-$\alpha$ lines of hydrogen isotopes in tokamaks \cite{Koubiti2022,Koubiti2023,Samuell2021,Saura2025}. 

In addition, we train models for Rosseland and Planck mean opacities as functions of temperature and density, for which a CNN alone yields highly accurate and computationally efficient predictions. The opacity spectra used in the training process are calculated with the {\sc Iliade} code \cite{Pain2006,Pain2021}. For both elements, large datasets spanning a wide range of temperatures and densities are used to train the models, with inputs given either as opacity spectra (functions of photon energy) or as scalar mean values. In practice, a temperature and a density are specified, and the model predicts either an opacity spectrum or a mean (Rosseland or Planck) opacity.
\\
\indent In Section \ref{sec:Opacity}, we briefly describe the {\sc Iliade} code which calculates the opacity in the superconfiguration approximation. A wide range of temperatures and densities is investigated, yielding opacity data for iron and nickel over energies up to 10 000 eV, with particular emphasis on the K-, L- and M-shells. In Section~\ref{sec:Deep learning}, we first examine the MLP model and then introduce a hybrid deep-learning architecture that integrates an MLP with a convolutional neural network and fully connected layers. The resulting CNN–MLP model substantially improves opacity predictions. The limitations of both approaches are discussed. In Section~\ref{sec:Predictions}, we benchmark the MLP and CNN--MLP predictions against selected {\sc Iliade} cases. Additionally, a standalone CNN model is trained to predict Rosseland and Planck mean opacities. Excellent agreement with direct calculations is obtained over a wide range of plasma conditions, spanning large temperature and density intervals. Concise descriptions of the MLP and the hybrid CNN–MLP model are presented in Appendix \ref{MLP} and Appendix \ref{CNN-MLP-FILM}, respectively.

\section{Opacity calculations}\label{sec:Opacity}
The spectral opacity $\kappa(E)$, where $E$ is the photon energy, includes contributions from bound-bound, bound-free and free-free transitions, as well as photon-electron scattering. Throughout this section, the dependence of the opacity on temperature and density is implicit in order to alleviate the notation:

\begin{equation*}
\kappa(E)=\kappa_{bb}(E)+\kappa_{bf}(E)+\kappa_{ff}(E)+\kappa_{\mathrm{scat}}(E).
\end{equation*}
Each contribution depends on both the atomic structure and the plasma state. Only the pure absorption coefficients are corrected for stimulated emission, as included in the corresponding $\kappa_{bb}$, $\kappa_{bf}$ and $\kappa_{ff}$ terms in the above equation. In general, robust atomic codes and reliable kinetic models are required to accurately compute atomic populations. Within the local thermodynamic equilibrium approximation, these calculations become straightforward.

\subsection{The superconfiguration approximation}
The {\sc Iliade} code is designed to compute the radiative opacity of hot plasmas. Such media, composed of multicharged ions in various excited states, exhibit an extremely large number of configurations, levels and transitions \cite{Krief2021}. Fortunately, when the density is high enough, individual line broadenings cause spectral lines to overlap, leading to broad collective features. In this regime, spectra can be efficiently modeled using statistical methods. 

One such approach is based on Unresolved Transition Arrays (UTAs) \cite{Bauche1979}, which approximate the ensemble of lines associated with a one-electron transition between two configurations by a Gaussian distribution. Its mean energy and variance can be obtained exactly through Racah algebra. However, at high temperatures and for high-$Z$ plasmas, the number of configurations (and consequently of UTAs) becomes prohibitively large.

The superconfiguration approximation addresses this issue by grouping configurations into classes called superconfigurations. A superconfiguration $\Xi$ is defined by a set of supershells (collections of subshells) and their electron populations. For instance
\[
\Xi = (1s\,2s\,2p)^{10} (3s\,3p\,3d)^{18} (4s\,4p\,4d\,4f\,5s\,5p\,5d)^{20},
\]
represents all configurations in which shells $n=1,2,3$ are fully occupied, while 20 electrons are distributed among the seven subshells from $4s$ to $5d$, consistently with the Pauli principle. The key idea is that the energy of any configuration $C$ within a given $\Xi$ can be approximated by a linear function of subshell populations:
\begin{equation}\label{eq7}
E_C \approx \sum_{s\in C} q_s \,\epsilon_s^{(\Xi)} 
+ \Big\langle E_{\tilde{C}} - \sum_{s'\in \tilde{C}} q_{s'} \,\epsilon_{s'}^{(\Xi)} \Big\rangle_{\tilde{C}\in\Xi},
\end{equation}
where $q_{s (s')}$ denotes the occupation number of subshell $s\,(s')$ belonging to $C\,(\tilde{C})$, and $\epsilon_{s\,(s')}^{(\Xi)}$ the one-electron eigenenergy common to all configurations in $\Xi$. The average, $\Big\langle \cdot \Big\rangle$, is performed over all the configurations $\tilde{C}$ belonging to $\Xi$. 

In the code, the calculation starts with an average-atom calculation, which gives the one-electron energies and wavefunctions at the plasma temperature and density. Then, in order to get realistic transition energies, these quantities are re-calculated for each superconfiguration. To this end, a self-consistent recalculation must be carried out using an approach similar to that employed in the average-atom model, but with integer electron populations. In the average-atom model, the electron populations are fractional numbers (averages over all the configurations of the plasma). Each superconfiguration is therefore characterized by a self-consistent potential defining its one-electron states.

This formalism makes it possible to describe the probabilities of states and configurations using the same expressions as in an ideal gas of independent electrons, provided that the energy spread of states in a superconfiguration is small compared with $k_BT$. All the superconfigurations are made of the same supershells (but with different populations), a supershell being an ensemble of subshells. This is what is meant by the ``supershell'' partition. There are different ways to group the subshells into supershells. Such a partition is obtained using a dedicated algorithm. This must not be confused with the ``partition functions'' of the supershells, which are also key ingredients of the superconfiguration formalism. The recursive nature of the method is its major advantage: convergence can be assessed by refining the supershell partition until the absorption spectrum becomes stable. 

The ensemble of transitions $a \to b$ between two superconfigurations is called a \emph{Supertransition Array} (STA) \cite{Barshalom1989}. Like UTAs, STAs are modeled statistically through their first three moments, with transition profiles represented by Gaussians. The $i$-th moment of transition $a \to b$ is:
\begin{equation*}
\mathcal{M}_i^{(\Xi,a\to b)} = 
\frac{4\pi^2\alpha}{3}
\sum_{C\in\Xi,\,C'\in\Xi'} \mathcal{P}_C 
\sum_{n\in C,\,m\in C'} (E_m - E_n)^i 
\left|\langle n|\hat{D}|m\rangle\right|^2,
\end{equation*}
where $\alpha$ is the fine-structure constant, $\langle n|\hat{D}|m\rangle$ the dipole matrix element, and $\mathcal{P}_C$ the temperature-dependent probability of configuration $C$. The approximation introduced in Eq.~(\ref{eq7}) makes this calculation tractable. This approach is still a subject of research, mainly aiming at controlling the resolution of the spectra \cite{Kurzweil2016}.

\subsection{Structure of the {\sc Iliade} code}
The computation begins with an average-atom calculation, which provides a self-consistent description of the plasma mean-electronic structure: orbital energies, wavefunctions, and fractional populations corresponding to the average configuration. The plasma is modeled within a Wigner–Seitz sphere, representing a fictitious ion whose charge equals the mean ionization. From this basis, {\sc Iliade} automatically generates a set of relevant superconfigurations. Supershells are constructed adaptively by grouping subshells close in energy. The ``divide-and-conquer'' algorithm \cite{Pain2021} recursively splits or merges supershells, ensuring an optimal partition without exceeding a user-defined maximum number of superconfigurations. For the retained superconfigurations, probabilities are computed and self-consistent-field calculations are performed, enforcing integer populations in each supershell via Lagrange multipliers. The resulting wavefunctions and free energies provide improved partition functions and transition energies. Relativistic corrections are included in the Pauli approximation, and exchange-correlation effects at finite temperature are treated using the Iyetomi–Ichimaru fit \cite{Iyetomi1986}. In addition, {\sc Iliade} accounts for configuration interaction between relativistic subconfigurations of a non-relativistic configuration. Several analytical models by Bar-Shalom and collaborators \cite{Barshalom1994,Barshalom2000} are implemented to correct transition intensities within STAs.

\subsection{Rosseland and Planck mean opacities}\label{mean opacities}
The Rosseland mean opacity $\kappa_R$ is defined as
\begin{equation}\label{moy-ros}
\frac{1}{\kappa_R} = 
\frac{\displaystyle \int_0^{\infty} \frac{1}{\kappa_\nu} \, 
\frac{\partial B_\nu(T)}{\partial T} \, d\nu}
{\displaystyle \int_0^{\infty} 
\frac{\partial B_\nu(T)}{\partial T} \, d\nu},
\end{equation}
where $\kappa_\nu$ is the frequency-dependent opacity, $T$ the plasma temperature, and $B_\nu$ the Planck function:
\begin{equation*}
B_\nu(T) = 
\frac{2h\nu^{3}}{c^{2}} \,
\frac{1}{e^{h\nu / k_{\mathrm{B}} T} - 1},
\end{equation*}
while the Planck mean opacity $\kappa_P$ is given by:

\begin{equation}\label{moy-pla}
\kappa_P=\frac{\displaystyle \int_0^{\infty} \kappa_\nu \, B_\nu(T)\, d\nu}
{\displaystyle \int_0^{\infty} B_\nu(T)\, d\nu}.
\end{equation}

\begin{figure}[htbp]
\centering
\includegraphics[scale=.6]{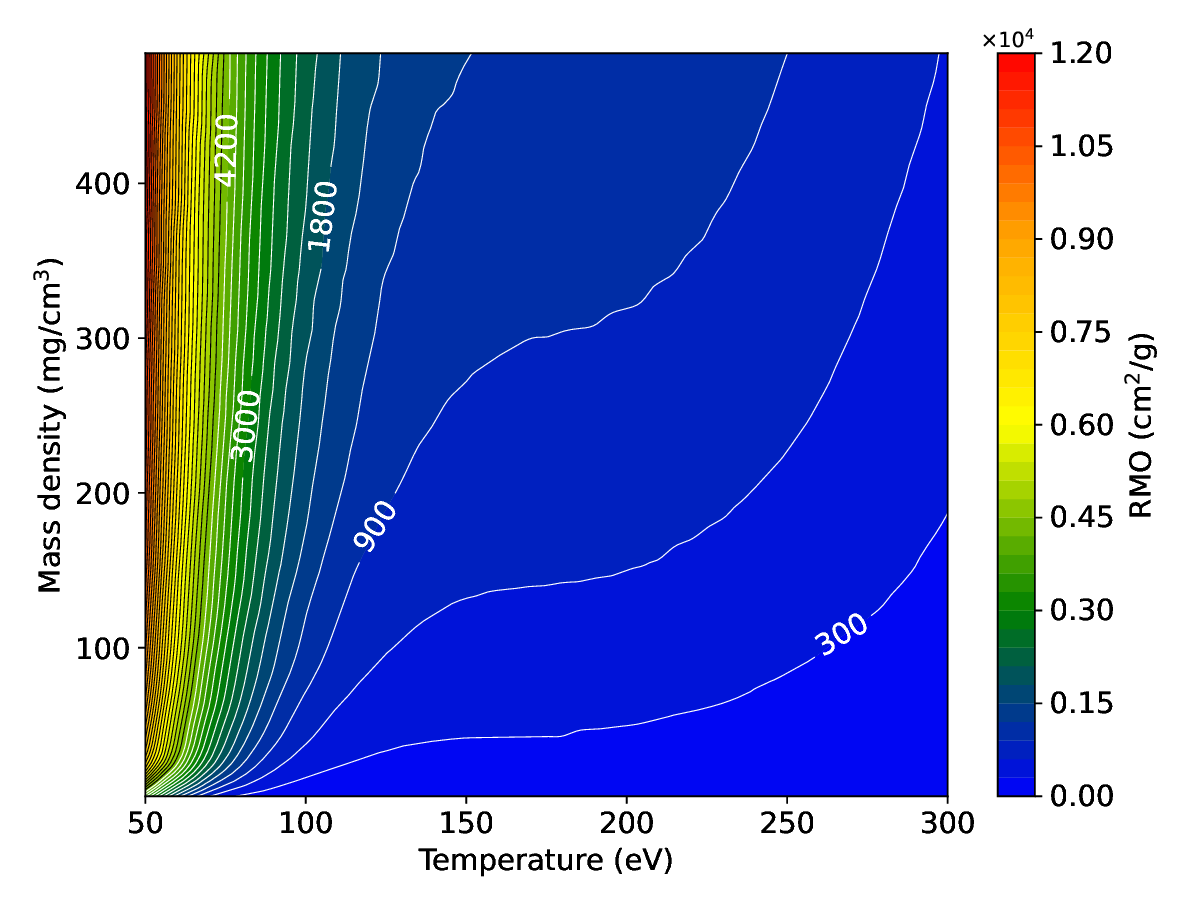}
\caption{Rosseland mean opacity (RMO) of iron as a function of temperature and mass density. The contour interval is 300 cm$^2$/g.}\label{Rosseland-Fe-Iliade}
\end{figure}

\begin{figure}[htbp]
\centering
\includegraphics[scale=.6]{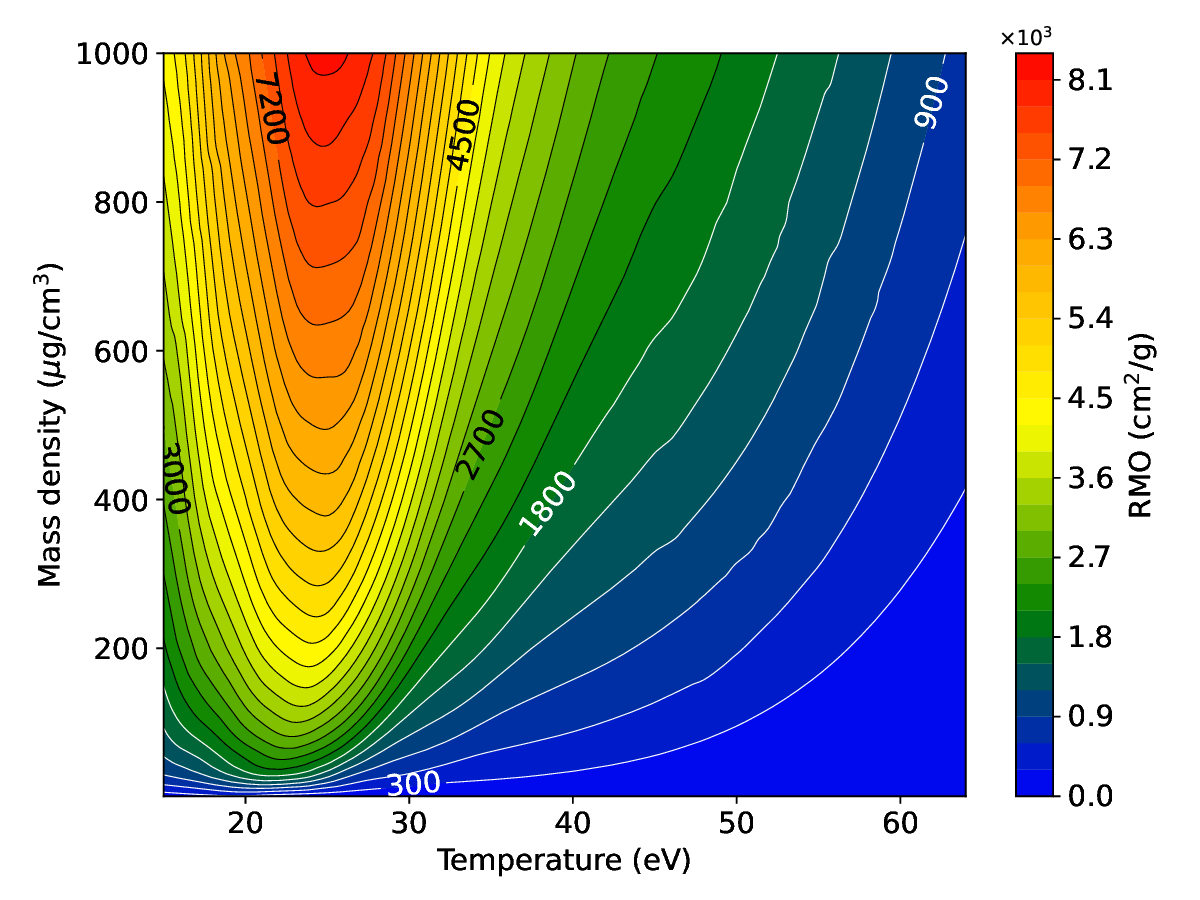}
\caption{Rosseland mean opacity (RMO) of nickel as a function of temperature and mass density. The contour interval is 300 cm$^2$/g.}\label{Rosseland-Ni-Iliade}
\end{figure}

\begin{figure}[htbp]
\centering
\includegraphics[scale=.6]{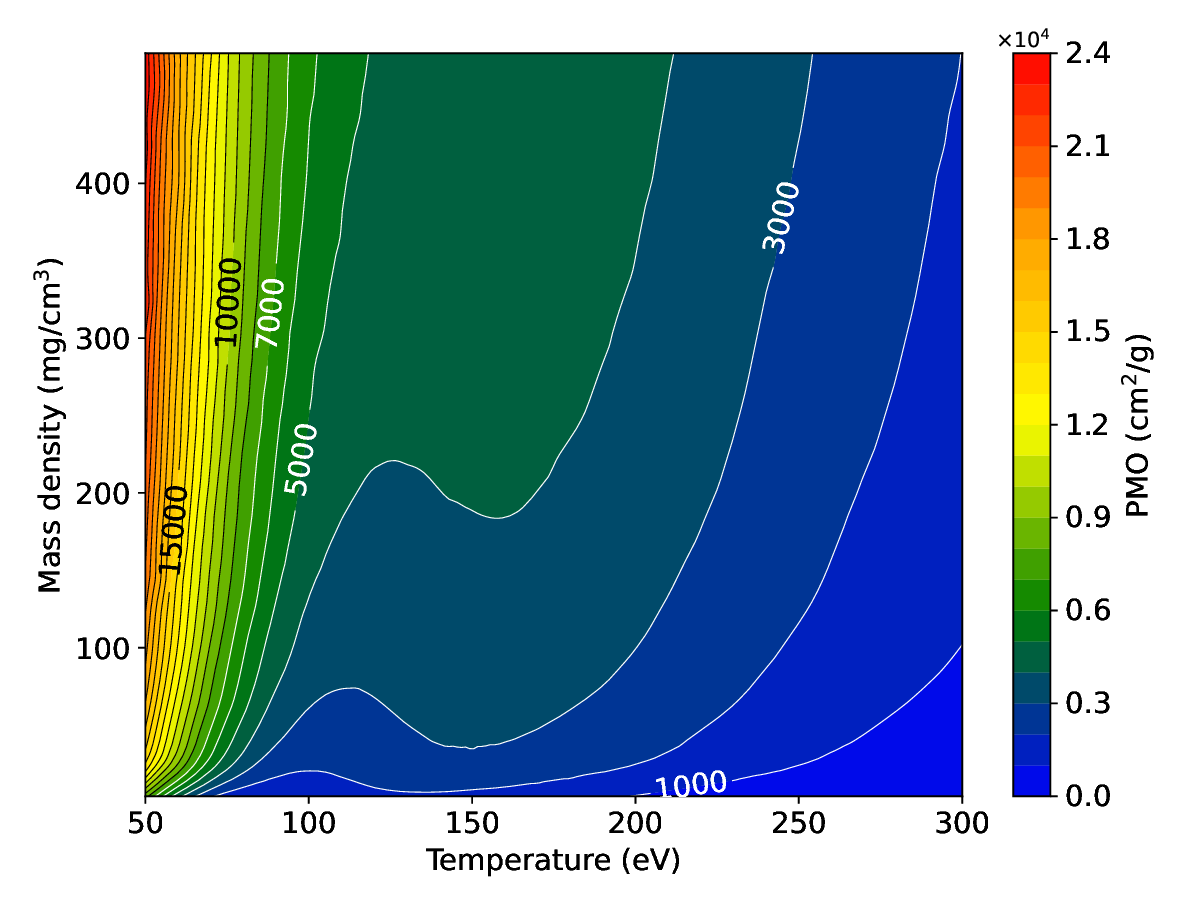}
\caption{Planck mean opacity (PMO) of iron as a function of temperature and mass density. The contour interval is 1000 cm$^2$/g.}\label{Planck-Fe-Iliade}
\end{figure}

\begin{figure}[htbp]
\centering
\includegraphics[scale=.6]{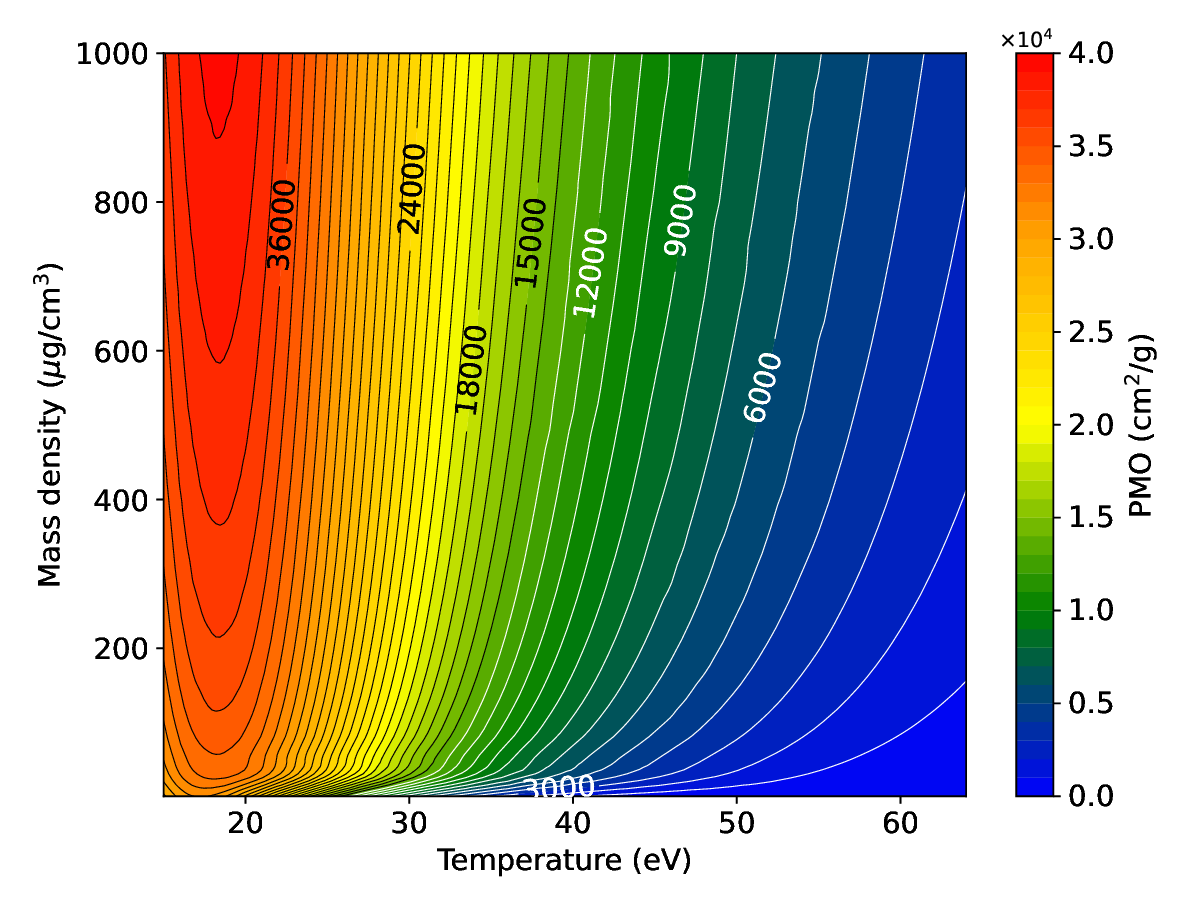}
\caption{Planck mean opacity (PMO) of nickel as a function of temperature and mass density. The contour interval is 1000 cm$^2$/g.}\label{Planck-Ni-Iliade}
\end{figure}

As the Rosseland mean is a harmonic mean, it is highly sensitive to spectral troughs —i.e., the gaps between lines—, and therefore to fine details of the absorption structure, in particular photoexcitation features. We calculated the mean opacities using opacity spectra generated with the {\sc Iliade} code. In Figs. \ref{Rosseland-Fe-Iliade}--\ref{Planck-Ni-Iliade}, we show how the mean opacities of iron and nickel vary with temperature and mass density. The thermodynamic conditions are not identical for iron and nickel, but in both cases the highest opacities are reached at relatively low temperatures and relatively high densities.

Conventional direct opacity calculations generally rely on atomic codes associated with substantial computational cost. In the following, we explore a deep-learning-based alternative aimed at significantly reducing calculation times.

\section{\label{sec:Deep learning}Deep learning}
Deep learning provides an efficient framework for predicting opacity spectra and for accelerating calculations in regimes where explicit physical models become computationally expensive \cite{Kluth2020, Schaeuble2025}.

In this work, we show that a hybrid CNN--MLP--FiLM architecture (see Appendix \ref{CNN-MLP-FILM}) is more effective than a standalone MLP (see Appendix \ref{MLP}) or a pure one-dimensional CNN to capture the functional relationship between the opacity $\kappa(E;T,\rho)$ and the photon energy $E=h\nu$, treated here as a quasi-continuous variable. The temperature and the mass density $\rho$ are treated as parameters. While an MLP alone treats each input energy point independently, the 1D CNN explicitly learns local and multiscale correlations along the energy grid, which is essential for accurately reproducing the shape of $\kappa$, particularly in the high-energy region. The thermodynamic parameters are incorporated through a Feature-wise Linear Modulation (FiLM) mechanism, in which an auxiliary MLP generates feature-dependent scaling and shifting coefficients that condition the CNN representations.\\
\indent Furthermore, we adopt a dichotomous training strategy by constructing two dedicated training procedures: one targeting the low-energy region (L- and M-shells) and the other the high-energy region (K-shell). This separation allows each model to specialize in the distinct statistical and physical characteristics of its respective energy range, thereby mitigating the dominance of low-energy data during training. As a result, the combined approach yields more accurate and robust predictions across the full energy spectrum, with an improvement in the description of high-energy spectral features.

\subsection{One-Dimensional CNN Encoder}
The One-Dimensional CNN encoder \cite{Churchill2020,Aymerich2022,Li2024,Kiranyaz2021} constitutes the core of our model for processing the sequential data associated with the variable $E$. It is well suited for ordered data in which local dependencies and position-specific patterns are meaningful. In our setting, the inputs to the CNN include not only the sequence of $E$ values, but also sequential representations of $T$ and $\rho$. Both plasma parameters are replicated along the length of the energy sequence. The input is then composed of three variables. To improve training stability, each variable $v$ is normalized as follows: $$\bar{v} = \frac{v - \mu}{\sigma},$$ where $\mu$ and $\sigma$ denote the mean and standard deviation of $v$, respectively, computed from the training data. After normalization, $\bar{v}$ has zero mean and unit variance. This multi-channel formulation allows the CNN to incorporate contextual information relating $\bar{E}$, $\bar{T}$, and $\bar{\rho}$ from the earliest stages of processing.

Convolution is a mathematical operation used to extract features from data (like edges in an image or patterns in sound). It involves sliding a small window of weights, called a kernel or filter, across the input data. At each position, it performs element-wise multiplication and sums the results to create a ``feature map'' that highlights specific characteristics. As can be seen in Appendix \ref{CNN-MLP-FILM}, CNN operates by applying convolutional filters that slide along the input sequence to extract local features. These filters compute responses that capture patterns such as gradients, peaks, and other localized variations. Each convolution is followed by Batch Normalization (BatchNorm), which normalizes the intermediate feature distributions to accelerate convergence, and improve training stability. BatchNorm is a technique used to make neural networks faster and more stable by recentering and rescaling the data. During training, the distribution of inputs to a layer can shift significantly; BatchNorm fixes this by normalizing the activations of a layer so they have a mean of 0 and a variance of 1 for each mini-batch. It prevents the network from getting stuck during training and allows for the use of higher learning rates. It is then followed by a non-linear activation function such as ReLU (Rectified Linear Unit), which is a mathematical formula applied to the output of a layer to introduce non-linearity into the network and allows the model to learn complex and non-linear relationships that cannot be captured by linear operations alone. Without this step, a neural network --no matter how many layers it has-- would behave like a simple linear model, and would be unable to learn complex patterns. The encoder is composed of a series of Residual Blocks, which enhance gradient flow and enable the training of deeper and more expressive architectures. Each block contains convolutional, normalization, and activation layers, together with a shortcut connection that adds the block’s input to its output. This residual pathway mitigates vanishing gradient issues and encourages the learning of residual mappings, which are generally easier to optimize than direct transformations. Overall, the encoder processes the sequential inputs $E$, $T$, and $\rho$ to extract hierarchical features and local structures. The resulting feature maps preserve the sequential nature of the data while encoding the essential relationships among the input variables, thereby providing a rich representation for predicting the spectral opacity.
\\
\indent In summary, the CNN encoder receives sequenced representations of $E$, $T$, and $\rho$, and employs convolutional layers and residual blocks to extract relevant features and relevant local patterns. Its outputs are enhanced sequential feature maps that encapsulate the key dependencies required for the prediction of $\kappa(E)$.

\subsection{MLP--FiLM conditioning}
The FiLM technique is a powerful method for dynamically conditioning the activations of a neural network based on plasma conditions. In our hybrid architecture, FiLM allows the model to modulate the output of the one-dimensional CNN encoder based on temperature and density values, which are inherently non-sequential conditions affecting the entire $\kappa(E)$ function.

At the heart of FiLM is a MLP. Unlike the CNN, which processes sequences, the MLP is designed to handle fixed-size inputs, in this case, the unique and normalized values of temperature and mass density for a given $\kappa(E)$ function. This MLP thus takes a two-element vector [$T$, $\rho$] as input and transforms it through several linear layers, each followed by a non-linear activation function (like ReLU). The output of the MLP is not the opacity $\kappa(E)$ itself. Instead, two vectors, $\boldsymbol{\gamma}(T,\rho)$ and $\boldsymbol{\beta}(T,\rho) \in \mathbb{R}^{C}$, are generated, where $C$ denotes the number of CNN feature channels. These vectors modulate the CNN features in a channel-wise manner according to
\begin{equation*}
{\rm FiLM}(h)=\boldsymbol{\gamma}(T,\rho)\odot h +\boldsymbol{\beta}(T,\rho),
\end{equation*}
where $h$ denotes an intermediate convolutional activation and $\odot$ the element-wise product. The above equation is applied independently to each channel $c$ ($c\in \lbrace 1, \cdots, C\rbrace$).

Conceptually, the hybrid architecture separates complementary tasks between its components: the CNN extracts and represents spectral features in the energy domain, while the MLP–FiLM pathway learns how these features evolve with the plasma conditions $(T,\rho)$. In this framework, FiLM plays a central role by conditioning intermediate CNN representations through affine transformations driven by the thermodynamic variables, thereby coupling spectral structure with plasma conditions in a controlled and physically consistent manner. This modulation enables the same learned spectral features to adapt smoothly and nonlinearly across different physical regimes, providing both flexibility and improved physical interpretability in the prediction of spectral opacity.

The training dataset involves 2000 nickel opacity spectra spanning temperatures of 15–64 eV and densities of 1–1000 $\mu$g/cm$^3$. For iron, the database includes 1275 opacity spectra with temperatures in the range 50–300 eV and densities of 4–484 mg/cm$^3$. Each spectrum involves $\sim 10^4$ energy values.

During the training phase, two files are generated: \textit{scaler.pkl} and \textit{modele.pth}. The former stores the preprocessing object used to normalize or standardize the input data before they are provided to the model. This ensures that the same transformations applied during training are consistently applied during inference, thereby maintaining compatibility between training and prediction data distributions.

The file \textit{modele.pth} contains the parameters learned by the neural network during training, including the weights and biases of the model (see Appendix \ref{MLP}). During the prediction phase, new input data are first transformed using the saved scaler and subsequently passed through the trained model loaded from \textit{modele.pth} to generate predictions.

\section{\label{sec:Predictions}Predictions}
\subsection{Opacity spectra}
We have used the training output to predict opacity spectra for iron and nickel in large intervals of temperatures and mass densities, relevant for laboratory plasmas as well as stellar plasmas. While the training necessitates a significant calculation time (tens of minutes at most, for opacity spectra, and around two minutes for Rosseland and Planck mean), the prediction itself is much faster. The principal advantage of our study lies in its computational efficiency and flexibility. Compared to standard LTE opacity tables with interpolation, the neural networks provide continuous functional representations of $\kappa(E;T,\rho)$ that can be evaluated significantly faster than repeated table lookups combined with high-resolution interpolation. In addition, the model reduces memory requirements by replacing large multi-dimensional tables with a compact set of learned parameters. We do not claim improved physical accuracy over the reference opacity model, but rather an efficient surrogate that preserves accuracy while enabling large-scale simulations and repeated evaluations.

\begin{figure}[htbp]
\centering
\newcommand{\HBracket}[8][]{%
  \pgfmathsetmacro{\xL}{#2} 
  \pgfmathsetmacro{\xR}{#3} 
  \pgfmathsetmacro{\yB}{#4} 
  \pgfmathsetmacro{\hL}{#5} 
  \pgfmathsetmacro{\hR}{#6} 
  \draw[#1] 
    (\xL,\yB) -- (\xL,\yB+\hL) -- (\xR,\yB+\hR) -- (\xR,\yB);
  \node[above, #1] at ({(\xL+\xR)/2},{\yB + max(\hL,\hR)}) {#7};
}

\begin{tikzpicture}
  \node[anchor=south west, inner sep=0] (img)
    at (0,0) {\includegraphics[width=0.8\linewidth]{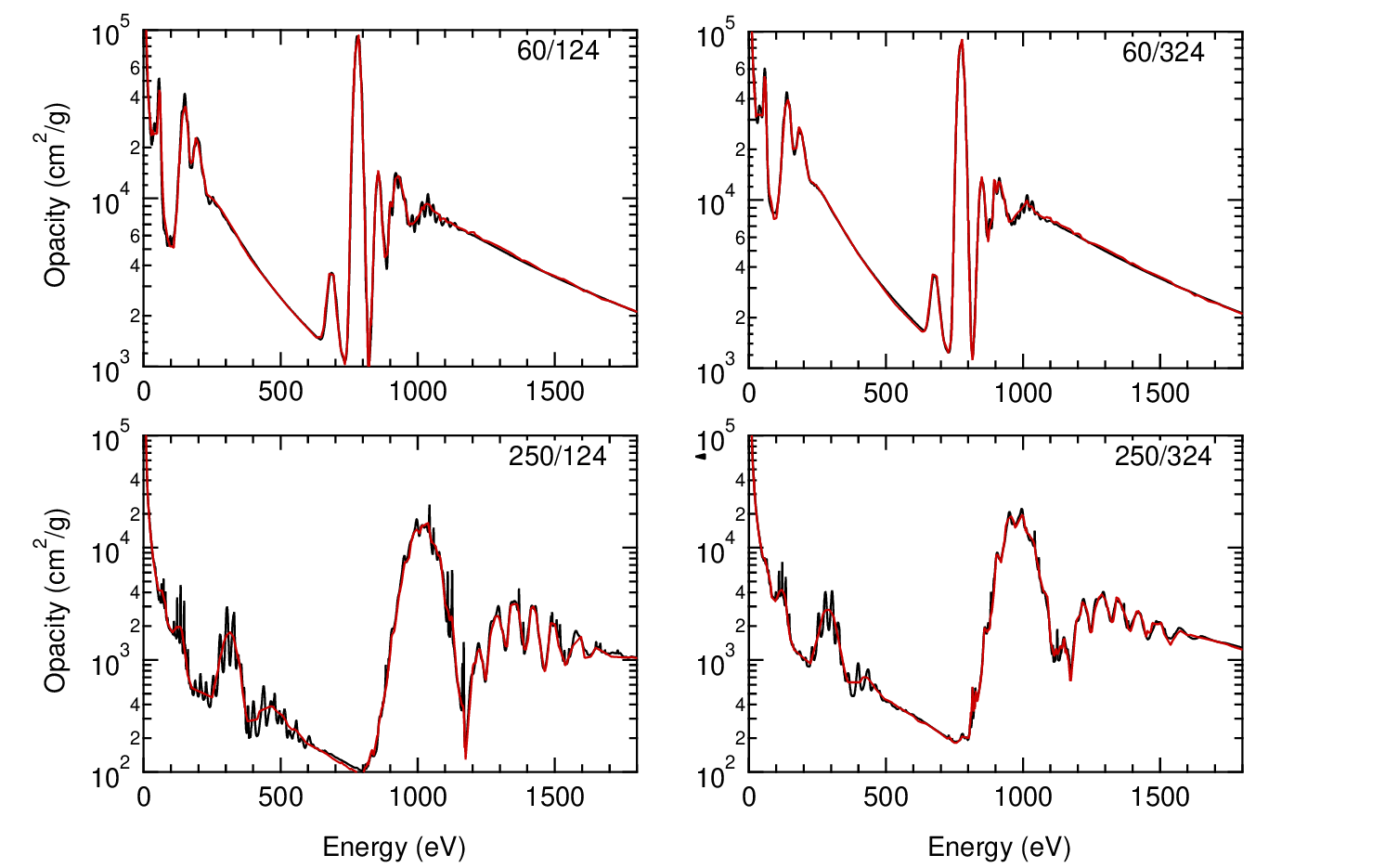}};

  \begin{scope}[x={(img.south east)}, y={(img.north west)}]

    \HBracket[thick, blue]{0.26}{0.35}{0.38}{0.02}{0.02}{\small 2p-3d (2s-3p,2p-3s)}{}

\node[
  font=\small,
  text=blue,
  anchor=west
] (ann1) at (0.15,0.91)
{3p--3d};

\draw[->, thick, blue]
  (ann1.west) -- (0.115,0.91);

\node[
  font=\small,
  text=blue,
  anchor=west
] (ann2) at (0.17,0.86)
{3p--4s};

\draw[->, thick, blue]
  (ann2.west) -- (0.135,0.86);

\node[
  font=\small,
  text=blue,
  anchor=west
] (ann3) at (0.71 ,0.88)
{2p--3d (2s--3p)};

\draw[->, thick, blue]
  (ann3.west) -- (0.7,0.86);

\node[
  font=\small,
  text=blue,
  anchor=south
] (ann4) at (0.65,0.75)
{2p--3s};

\draw[->, thick, blue]
  (ann4.south) -- (0.675,0.69);
    
  \end{scope}
\end{tikzpicture}
\caption{M- and L-shell opacity of iron as a function of photon energy. In each panel, a/b indicates the temperature (eV) and mass density (mg/cm$^3$). Black curves correspond to calculations performed with the {\sc Iliade} code, while red curves show MLP predictions.}\label{OPA_Fe_low_2}
\end{figure}
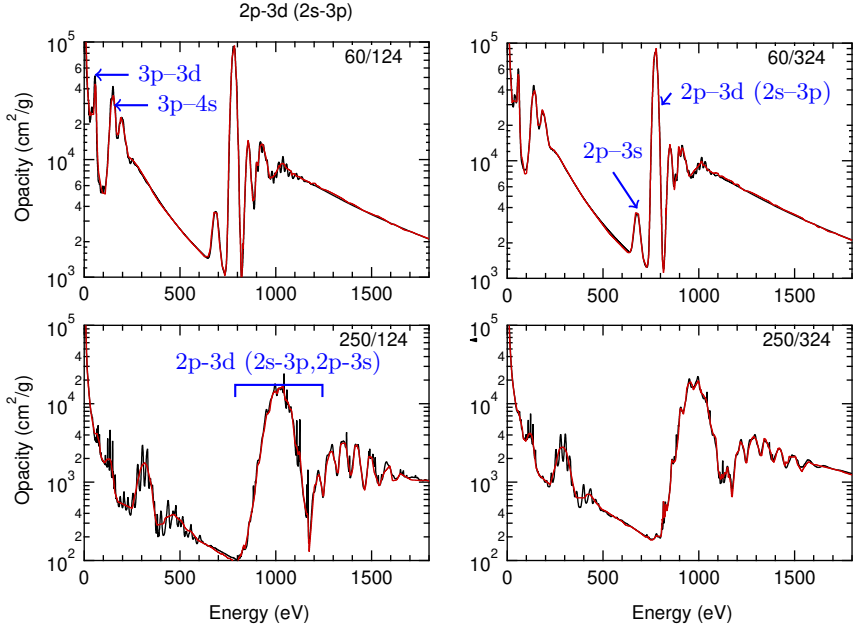

In Fig. \ref{OPA_Fe_low_2}, we compare calculated and predicted low-energy iron opacities for two temperatures (60 and 250 eV) and two densities (124 and 324 mg/cm$^ 3$). The predicted spectra are obtained using a multilayer perceptron (see Appendix \ref{MLP}). The model is trained by iteratively adjusting the interlayer weights via forward propagation and backpropagation so as to minimize the prediction error, enabling the training of complex nonlinear relationships. Training is performed over 100 epochs. An epoch is one complete pass of the entire training dataset through the model. At low photon energies, the MLP provides accurate predictions and shows good agreement with the reference {\sc Iliade} profiles. 

At 60 eV, the Boltzmann factor $\exp(-\Delta E/60)$ decreases rapidly with increasing transition energy $\Delta E$, such that levels with principal quantum numbers $n=4,5,6$ lie far above the characteristic thermal energy and remain only weakly populated. As a result, the opacity spectrum is dominated by the lowest energy 2p-3d transition array. As the temperature increases from 60 to 250 eV, the number of contributing spectral lines increases substantially, leading to the progressive merging of the 2p-3d, 2s-3p, and 2p-3s transition arrays. This behavior is accompanied by a significant increase in the population of excited states. At 250 eV, the thermal energy $k_{\mathrm{B}}T$ becomes comparable to, or exceeds, the excitation energy required to access $n\geq 4$ shells. Consequently, the Boltzmann factor increases significantly, leading to substantial population of higher-lying levels. Under local thermodynamic equilibrium conditions, the plasma also shifts toward much higher ionization stages. For a given ion, increased ionization means higher ionization potential; however, the 2p-nd transitions observed at 250 eV often originate from ions that are not present at 60 eV. These new, more highly charged, ions have electronic structures for which transitions to $n\geq 4$ levels are either stronger or more clearly resolved from neighboring spectral features. As expected, increasing the temperature from 60 to 250 eV results in markedly more complex spectral structure, which is well reproduced in the present case. As shown in the upper-right panel of Fig. \ref{OPA_Fe_low_2}, the 2p–3d and 2s–3p transition arrays merge, with the 2p–3d contribution remaining dominant. In contrast, in the lower-left panel, the 2s–3p and 2p–3s transition arrays (indicated in parentheses) are weaker than the 2p–3d transition array.

\begin{figure}[htpb]
\centering
\newcommand{\HBracket}[8][]{%
  \pgfmathsetmacro{\xL}{#2} 
  \pgfmathsetmacro{\xR}{#3} 
  \pgfmathsetmacro{\yB}{#4} 
  \pgfmathsetmacro{\hL}{#5} 
  \pgfmathsetmacro{\hR}{#6} 
  \draw[#1] 
    (\xL,\yB) -- (\xL,\yB+\hL) -- (\xR,\yB+\hR) -- (\xR,\yB);
  \node[above, #1] at ({(\xL+\xR)/2},{\yB + max(\hL,\hR)}) {#7};
}

\begin{tikzpicture}
  \node[anchor=south west, inner sep=0] (img)
    at (0,0) {\includegraphics[width=0.8\linewidth]{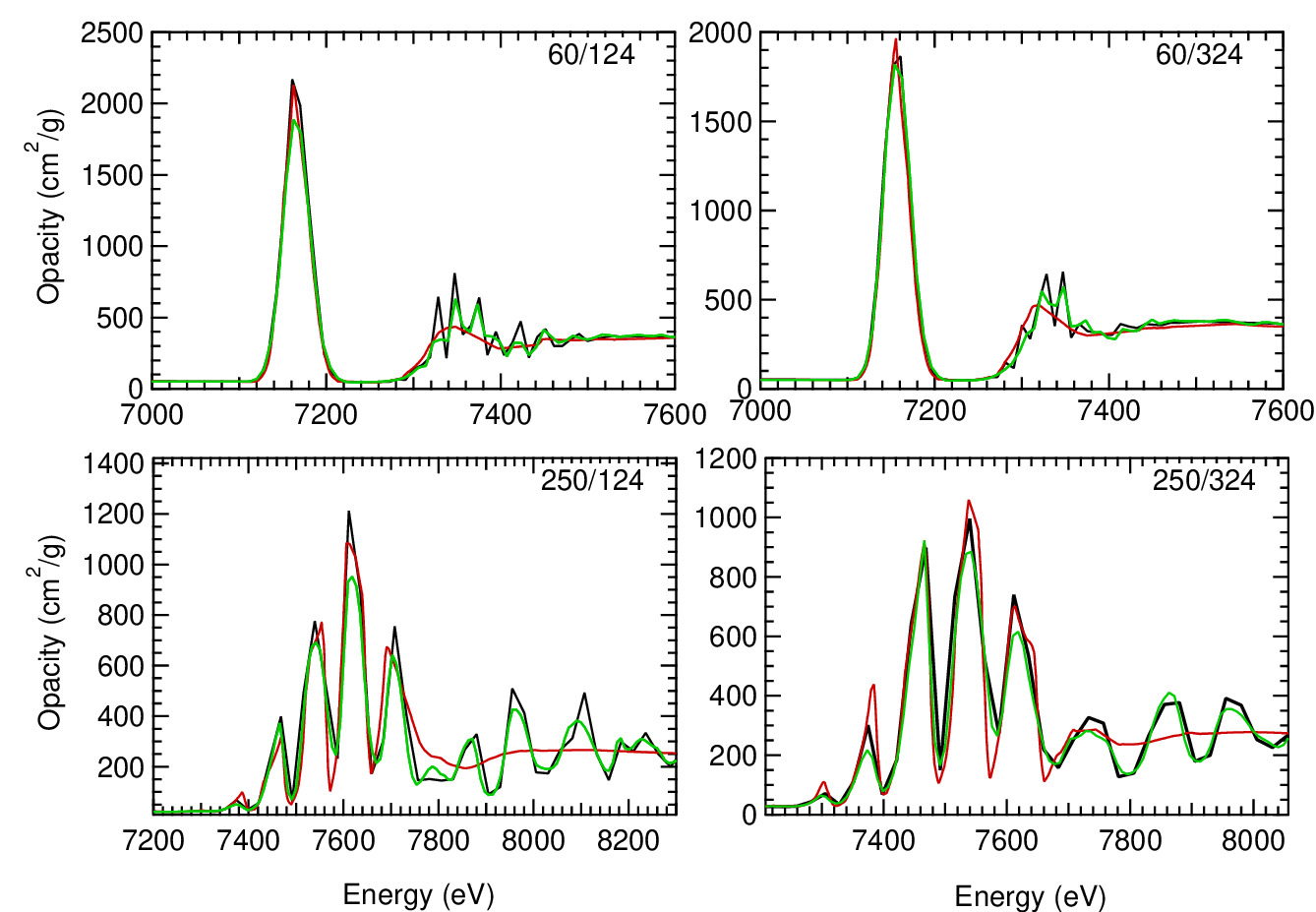}};

  \begin{scope}[x={(img.south east)}, y={(img.north west)}]

    \HBracket[thick, blue]{0.32}{0.43}{0.685}{0.02}{0.02}{\small 1s-np}{}
    \HBracket[thick, blue]{0.16}{0.5}{0.4}{0.03}{0.03}{\small 1s-mp}{}

\node[
  font=\small,
  text=blue,
  anchor=west
] (ann1) at (0.27,0.83)
{1s--4p};

\draw[->, thick, blue]
  (ann1.west) -- (0.23,0.8);
   
  \end{scope}
\end{tikzpicture}
\caption{K-shell opacity of iron as a function of photon energy. In each panel, a/b denotes the temperature (eV) and mass density (mg/cm$^3$). The black, red, and green curves correspond to calculations with the {\sc Iliade} code, MLP, and CNN-MLP-FiLM predictions, respectively. In the upper-left panel $n$ ranges from 5 to 7, while in the lower-left panel $m$ ranges from 3 to 9.}\label{OPA_Fe_high_2}
\end{figure}

At high energies (K-shell), MLP predictions show poor agreement with the {\sc Iliade} opacity profiles, with several spectral peaks missing, as shown in Fig. \ref{OPA_Fe_high_2}. In fact, MLP-based predictions fail to reproduce many features in the 7300--8300 eV energy range. We therefore adopt the hybrid (CNN–MLP-FiLM) approach described above and illustrated in Appendix \ref{CNN-MLP-FILM}, which yields substantially improved results in all cases. Incorporating the CNN encoder markedly enhances the agreement with {\sc Iliade} calculations by accurately capturing and localizing all spectral peaks. Increasing the temperature from 60 to 250 eV suppresses the 1s--4p peak owing to a strong reduction in the $n=4$ population, while peaks associated with $n>4$ increase by approximately a factor $\simeq 2$.

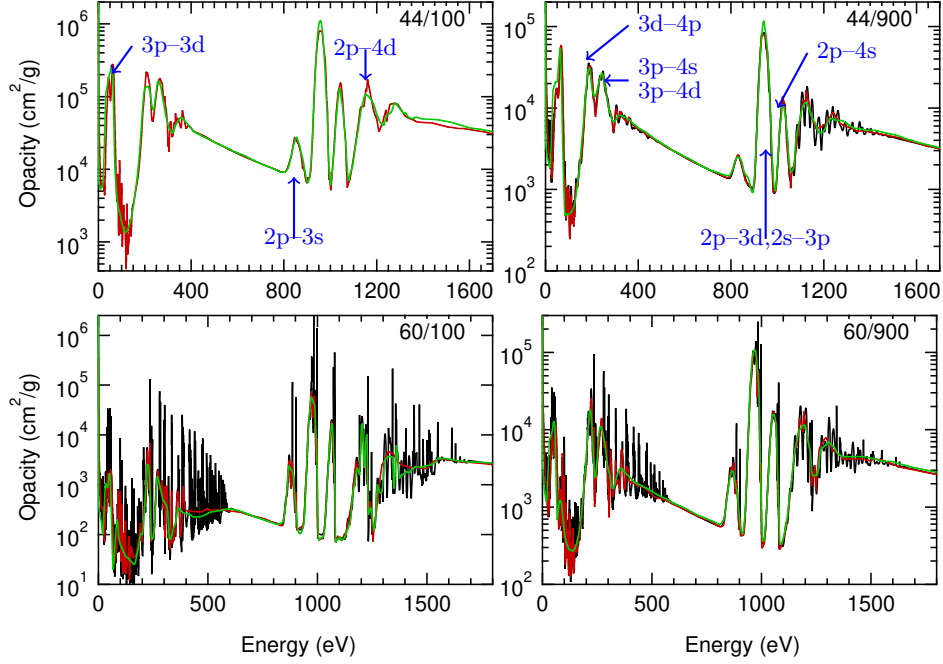
\begin{figure}[htbp]
\centering

\newcommand{\HBracket}[8][]{%
  \pgfmathsetmacro{\xL}{#2} 
  \pgfmathsetmacro{\xR}{#3} 
  \pgfmathsetmacro{\yB}{#4} 
  \pgfmathsetmacro{\hL}{#5} 
  \pgfmathsetmacro{\hR}{#6} 
  \draw[#1] 
    (\xL,\yB) -- (\xL,\yB+\hL) -- (\xR,\yB+\hR) -- (\xR,\yB);
  \node[above, #1] at ({(\xL+\xR)/2},{\yB + max(\hL,\hR)}) {#7};
}

\begin{tikzpicture}
  \node[anchor=south west, inner sep=0] (img)
    at (0,0) {\includegraphics[width=0.82\linewidth]{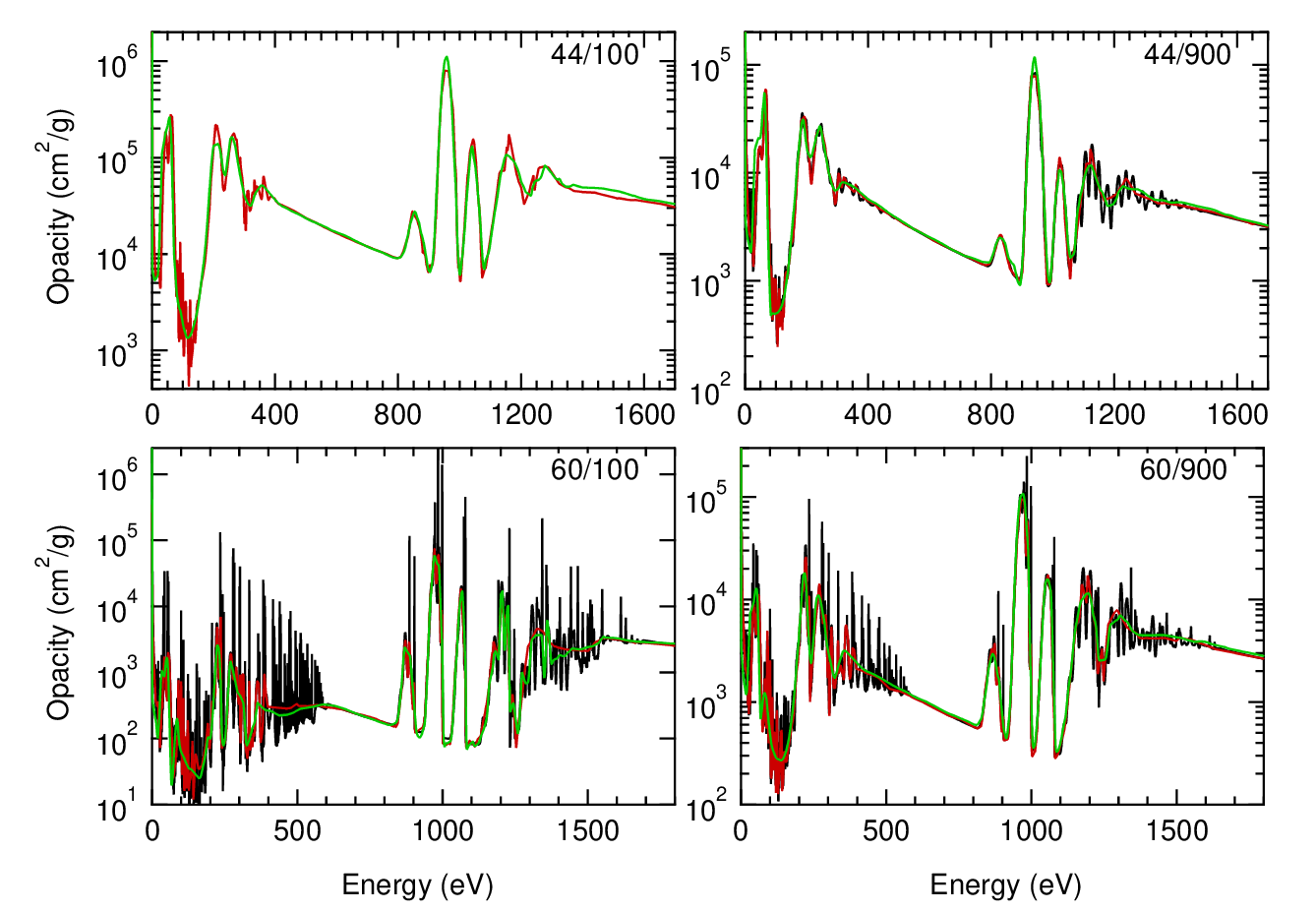}};

  \begin{scope}[x={(img.south east)}, y={(img.north west)}]

\node[
  font=\small,
  text=blue,
  anchor=west
] (ann2) at (0.15,0.9)
{3p--3d};

\draw[->, thick, blue]
  (ann2.west) -- (0.13,0.86);

\node[
  font=\small,
  text=blue,
  anchor=west
] (ann2) at (0.83,0.89)
{2p--4s};

\draw[->, thick, blue]
  (ann2.west) -- (0.8,0.81);

\node[
  font=\small,
  text=blue,
  anchor=center
] (ann2) at (0.385,0.895)
{2p--4d};

\draw[->, thick, blue]
  (ann2.center) -- (0.385,0.85);
  
\node[
  font=\small,
  text=blue,
  anchor=center
] (ann3) at (0.313,0.62)
{2p--3s};

\draw[->, thick, blue]
  (ann3.center) -- (0.313,0.71);

\node[
  font=\small,
  text=blue,
  anchor=center
] (ann2) at (0.788,0.62)
{2p--3d,2s--3p};

\draw[->, thick, blue]
  (ann2.center) -- (0.788,0.75);

\node[
  font=\small,
  text=blue,
  anchor=west
] (ann3) at (0.65,0.93)
{3d--4p};

\draw[->, thick, blue]
  (ann3.west) -- (0.607,0.88);

\node[
  font=\small,
  text=blue,
  anchor=west
] (ann4) at (0.65,0.85)
{\shortstack[l]{3p--4s\\[-2pt]3p--4d}};

\draw[->, thick, blue]
  (ann4.west) -- (0.625,0.85);
    
  \end{scope}
\end{tikzpicture}
\caption{M- and L-shell opacity of nickel as a function of photon energy. In each panel, a/b denotes the temperature (eV) and mass density ($\mu$g/cm$^3$). The black, red, and green curves correspond to calculations with {\sc Iliade} code, MLP and CNN-MLP-FiLM predictions, respectively.}\label{OPA_Ni_low_2}
\end{figure}

We now present opacity predictions for nickel. The temperature and mass density values used in the training span the ranges 15--64 eV and 1--1000 $\mu$g/cm$^3$, respectively. In the following, we assess the prediction accuracy for plasmas at temperatures of $T$=44 and 60 eV and mass densities of 100 and 900 $\mu$g/cm$^3$. In the low-energy region, MLP and CNN–MLP-FiLM produce similar and satisfactory results at 44 eV (see Figure \ref{OPA_Ni_low_2}, upper panels). At $T$=60 eV (lower panels), the rapidly growing number of spectral peaks leads to less accurate, though still acceptable, predictions for both models. Training of the CNN–MLP-FiLM model takes approximately 2–3 times longer than that of the MLP model, mainly due to the substantially larger number of transitions in the M- and L-shells compared with the K-shell.

\begin{figure}[htpb]
\centering
\newcommand{\HBracket}[8][]{%
  \pgfmathsetmacro{\xL}{#2} 
  \pgfmathsetmacro{\xR}{#3} 
  \pgfmathsetmacro{\yB}{#4} 
  \pgfmathsetmacro{\hL}{#5} 
  \pgfmathsetmacro{\hR}{#6} 
  \draw[#1] 
    (\xL,\yB) -- (\xL,\yB+\hL) -- (\xR,\yB+\hR) -- (\xR,\yB);
    \node[above, #1] at ({(\xL+\xR)/2},{\yB + max(\hL,\hR)}) {#7};
}
\begin{tikzpicture}
  \node[anchor=south west, inner sep=0] (img)
    at (0,0) {\includegraphics[width=0.8\linewidth]{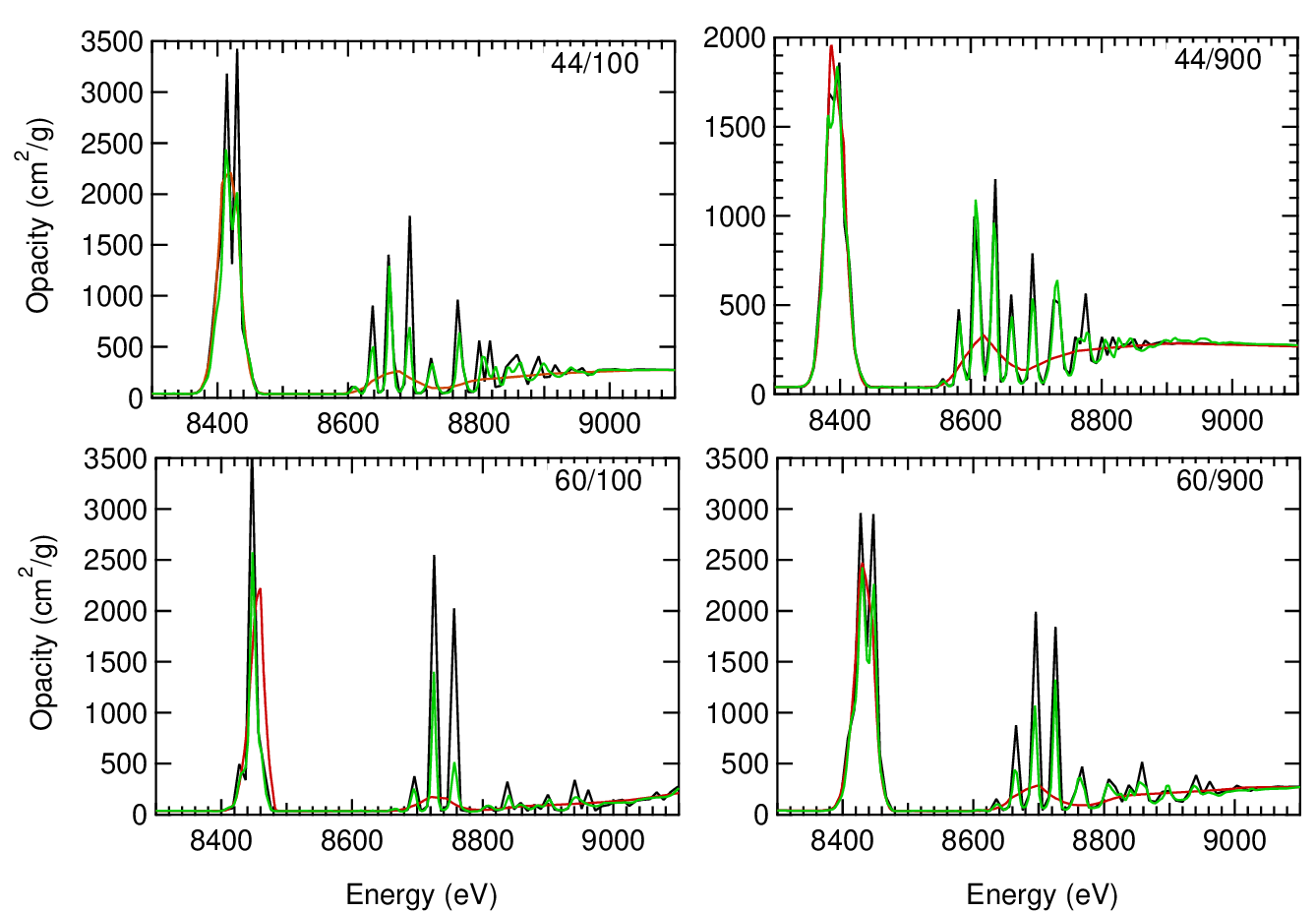}};

  \begin{scope}[x={(img.south east)}, y={(img.north west)}]

    \HBracket[thick, blue]{0.26}{0.35}{0.74}{0.02}{0.02}{\small 1s-mp}{}
    
    \node[
  font=\small,
  text=blue,
  anchor=west
] (ann2) at (0.23,0.85)
{1s--np};

\draw[->, thick, blue]
  (ann2.west) -- (0.182,0.8);

    \node[
  font=\small,
  text=blue,
  anchor=west
] (ann2) at (0.225,0.9)
{1s--4p};

\draw[->, thick, blue]
  (ann2.west) -- (0.17,0.9);

  \end{scope}
\end{tikzpicture}


\caption{K-shell opacity of nickel as a function of photon energy. In each panel, a/b indicates the temperature (eV) and mass density ($\mu$g/cm$^3$). The black, red, and green curves correspond to {\sc Iliade} calculations, MLP, and CNN-MLP-FiLM predictions, respectively. In the upper-left panel $n=5, 6$ and $m\geq 7$.}\label{OPA_Ni_high}
\end{figure}
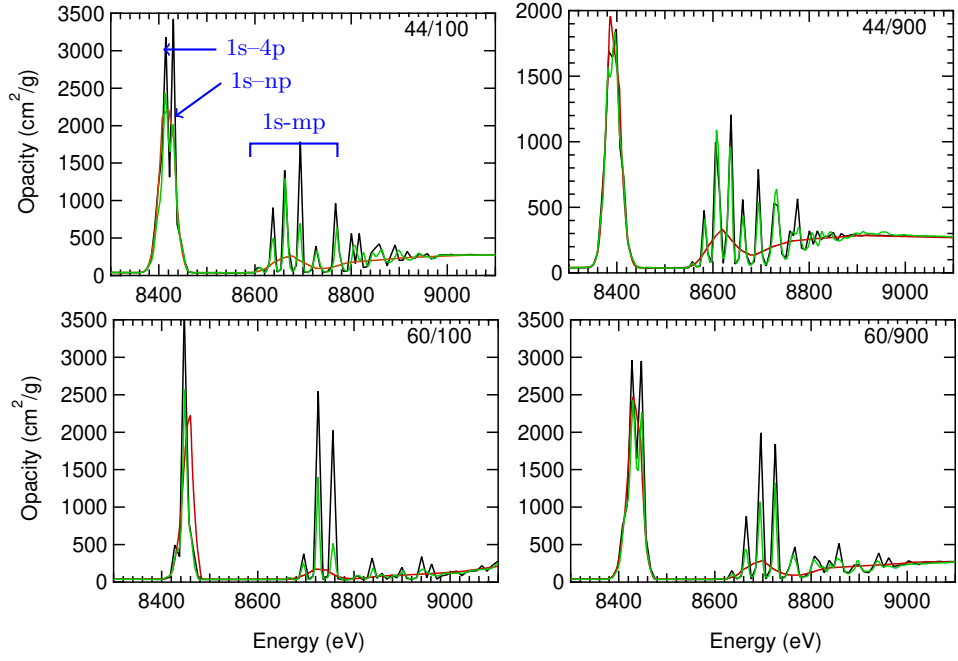

To interpret the evolution of the opacity spectrum with changing thermodynamic conditions, it is necessary to examine transitions involving the M-shell. At 44 eV, nickel is predominantly in moderately ionized states, typically around Ni IX--Ni X. By 60 eV, the $n=3$ shell becomes significantly depleted, giving rise to a sharp increase in combinatorial complexity. Nickel possesses a large number of electrons in the 3d subshell, and as these electrons are excited or ionized, the number of possible couplings of orbital and spin angular momenta ($L$ and $S$) increases dramatically. Even a single configuration, such as 3p$^5$3d$^2$, can generate several tens to hundreds of distinct energy levels, each serving as an initial or final state for additional spectral lines. A second important effect is the rapid increase of $\Delta n$=0 (3s-3p, 3p-3d) and $n=3$ to $n=4$ (3s-4p, 3p-4s, 3p-4d, 3d-4p, 3d-4f) transitions. Between 44 and 60 eV, transitions such as 3p-3d arise. These transitions are extremely numerous and very close in energy, leading to the formation of UTAs. At 44 eV, the thermal energy remains just below the population threshold for these levels, and such transitions are just emerging. At 60 eV, the threshold is crossed, and thousands of lines emerge simultaneously in the opacity spectrum.\\
\indent The charge-state distribution is also highly sensitive to temperature under these low-density conditions. An increase in temperature from 44 to 60 eV corresponds to a 36 \% rise in thermal energy, which appears within the exponential terms of the Boltzmann and Saha relations. Consequently, even modest temperature variations can result in a 10- or 100-fold increase in the population of excited states when their excitation energies lie near 50-60 eV. This sensitivity is further amplified at low densities. At $\rho \simeq 100\  \mu$g/cm$^3$, line broadening due to collisions or ion Stark effect is negligible. The lines remain narrow, allowing the progressive appearance of new spectral features to be clearly resolved. At higher densities ($\rho \simeq 900\ \mu$g/cm$^3$), spectral lines overlap and merge into smooth and broad features. The spectrum thus evolves from a sparse, comb-like structure into a dense and highly congested forest of lines. In summary, for nickel, the temperature range 44-60 eV represents a critical tipping point at which electrons in the 3d subshell begin to play a significant role. It is this change in electronic structure, rather than the increase in thermal energy alone, that drives the dramatic multiplication of spectral lines in the opacity spectrum.\\
\indent Finally, we present the high-energy results in Fig. \ref{OPA_Ni_high}. The hybrid model clearly outperforms the MLP model, accurately localizing all peaks, whereas the latter does not. Peak heights are predicted more accurately at a temperature of 44 eV than at 60 eV.

\begin{figure}[htpb]
\centering
\includegraphics[scale=.45]{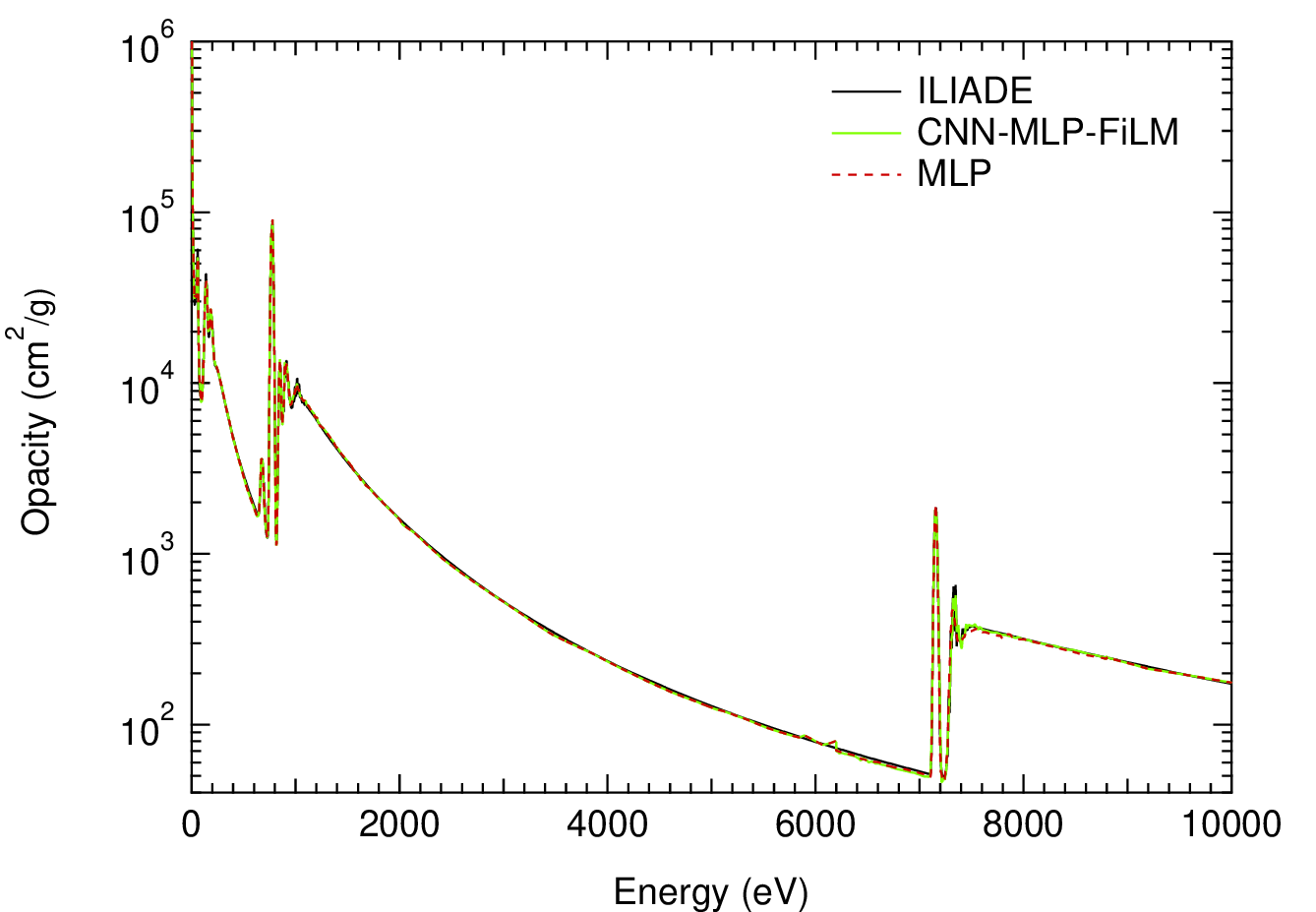}
\caption{Opacity profile of iron as a function of energy. Temperature: 60 eV; mass density: 324 mg/cm$^3$. Black line: Calculation with {\sc Iliade} code; red dashed line: MLP; green line: CNN-MLP-FiLM.}\label{OPA_Fe_60_324}
\end{figure}

\begin{figure}[htpb]
\centering
\includegraphics[scale=.45]{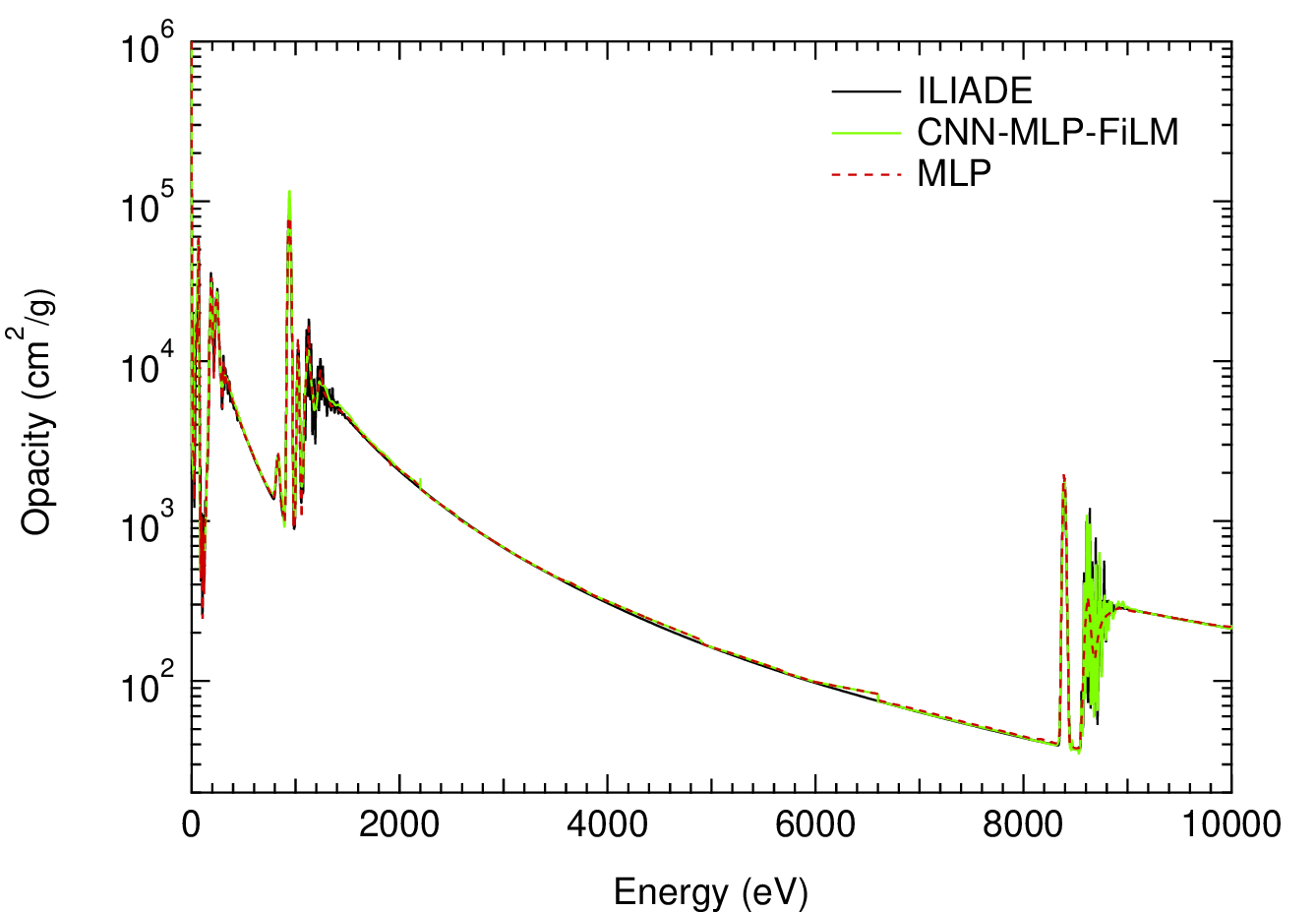}
\caption{Opacity profile of nickel as a function of energy. Temperature: 44 eV; mass density: 0.9 mg/cm$^3$. Black line: Calculation with {\sc Iliade} code; red dashed line: MLP; green line: CNN-MLP-FiLM.}\label{OPA_Ni_44_900}
\end{figure}

An overview of the opacity profiles across the entire energy range is shown in Fig. \ref{OPA_Fe_60_324} (iron) and Fig. \ref{OPA_Ni_44_900} (nickel). The predicted opacities show satisfactory agreement with {\sc Iliade} calculations.

In the following, we focus on the Rosseland and Planck mean opacities of iron and nickel as functions of density and temperature. Since we deal with scalars rather than full functions, the training is simpler and faster. Accordingly, we train a one-dimensional CNN model rather than a hybrid one.

\subsection{Mean-opacity predictions}
The second objective of this work is to predict the Rosseland and Planck mean opacities (Eqs.~(\ref{moy-ros}) and (\ref{moy-pla}), respectively). Unlike opacity spectra, which are evaluated on energy grids of order $10^4$ points, this task involves scalar outputs. We therefore train a one-dimensional convolutional neural network, which enables rapid training and yields highly accurate predictions for both $\kappa_R$ and $\kappa_P$. The temperature and density ranges involved in the learning process are identical to those in Section \ref{mean opacities}.

\begin{figure}[htpb]
\centering
\includegraphics[scale=.45]{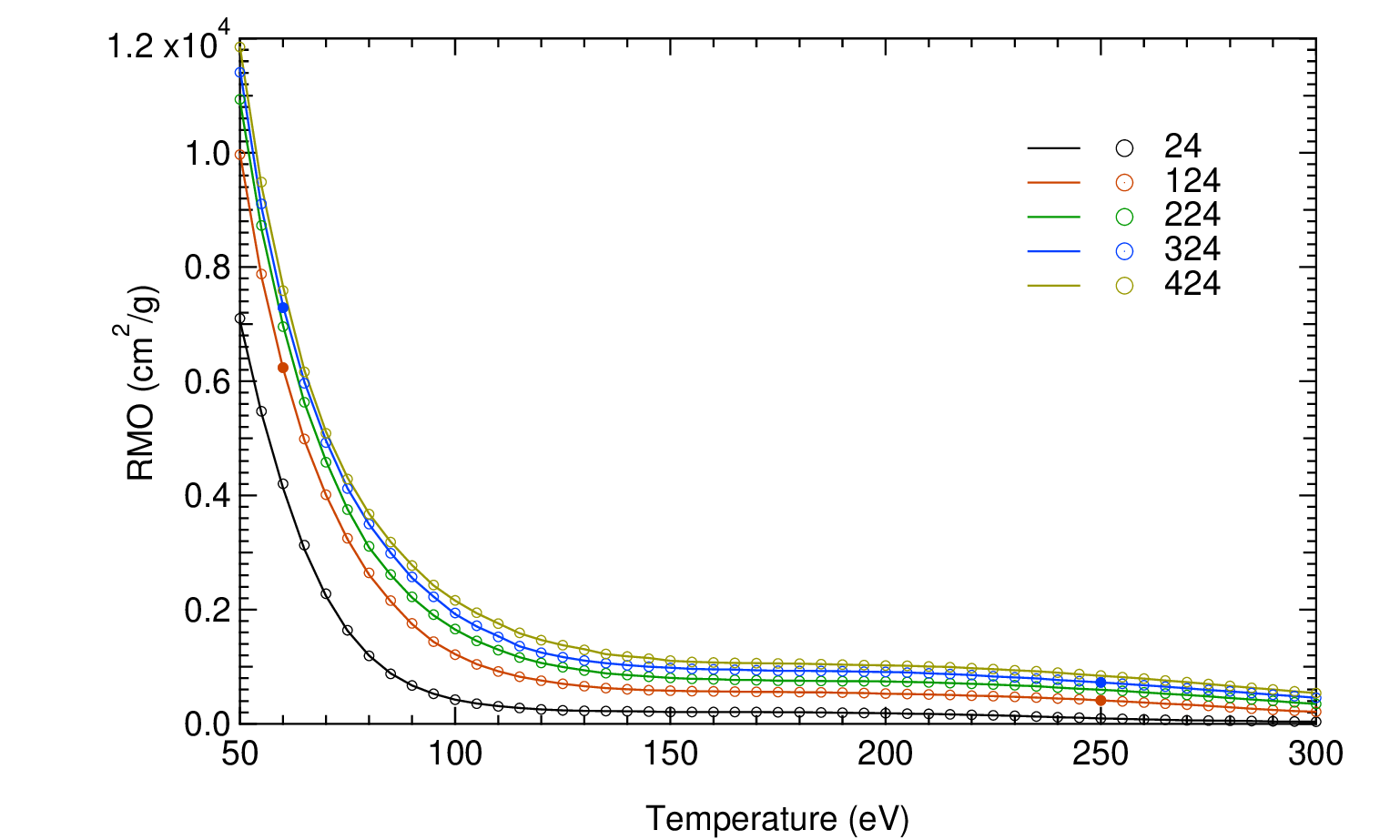}
\caption{Rosseland mean opacity of iron as a function of temperature for five mass densities (in mg/cm$^3$). Lines represent calculated values; open circles indicate predicted values; filled circles correspond to the cases shown in Figs. \ref{OPA_Fe_low_2} and \ref{OPA_Fe_high_2}.}\label{Pred_Fer_Rosseland}
\end{figure}

\begin{figure}[htpb]
\centering
\includegraphics[scale=.45]{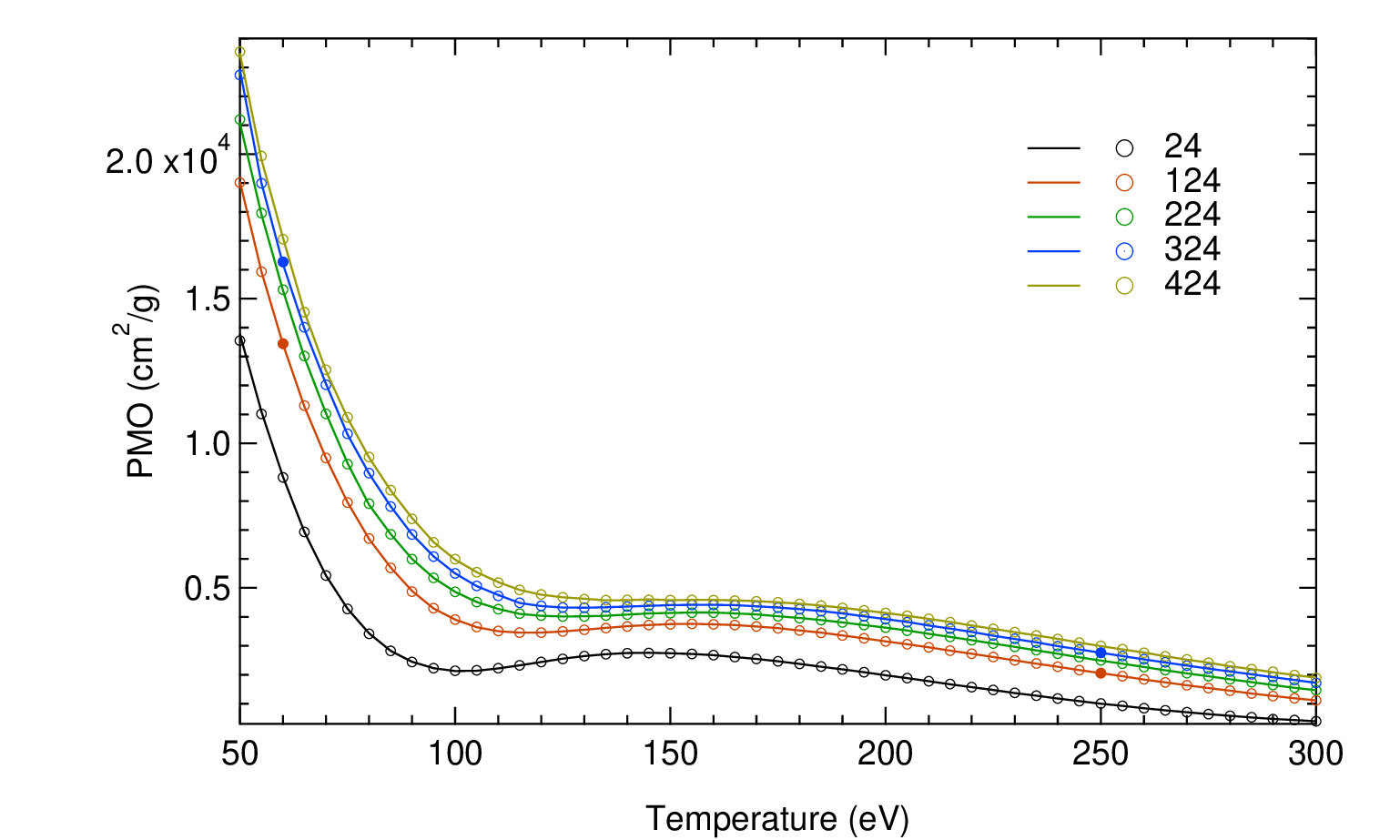}
\caption{Planck mean opacity of iron as a function of temperature for five mass densities (in mg/cm$^3$). Lines represent calculated values; open circles indicate predicted values; filled circles correspond to the cases shown in Figs. \ref{OPA_Fe_low_2} and \ref{OPA_Fe_high_2}.}\label{Pred_Fer_Planck}
\end{figure}

\begin{figure}[htpb]
\centering
\includegraphics[scale=.45]{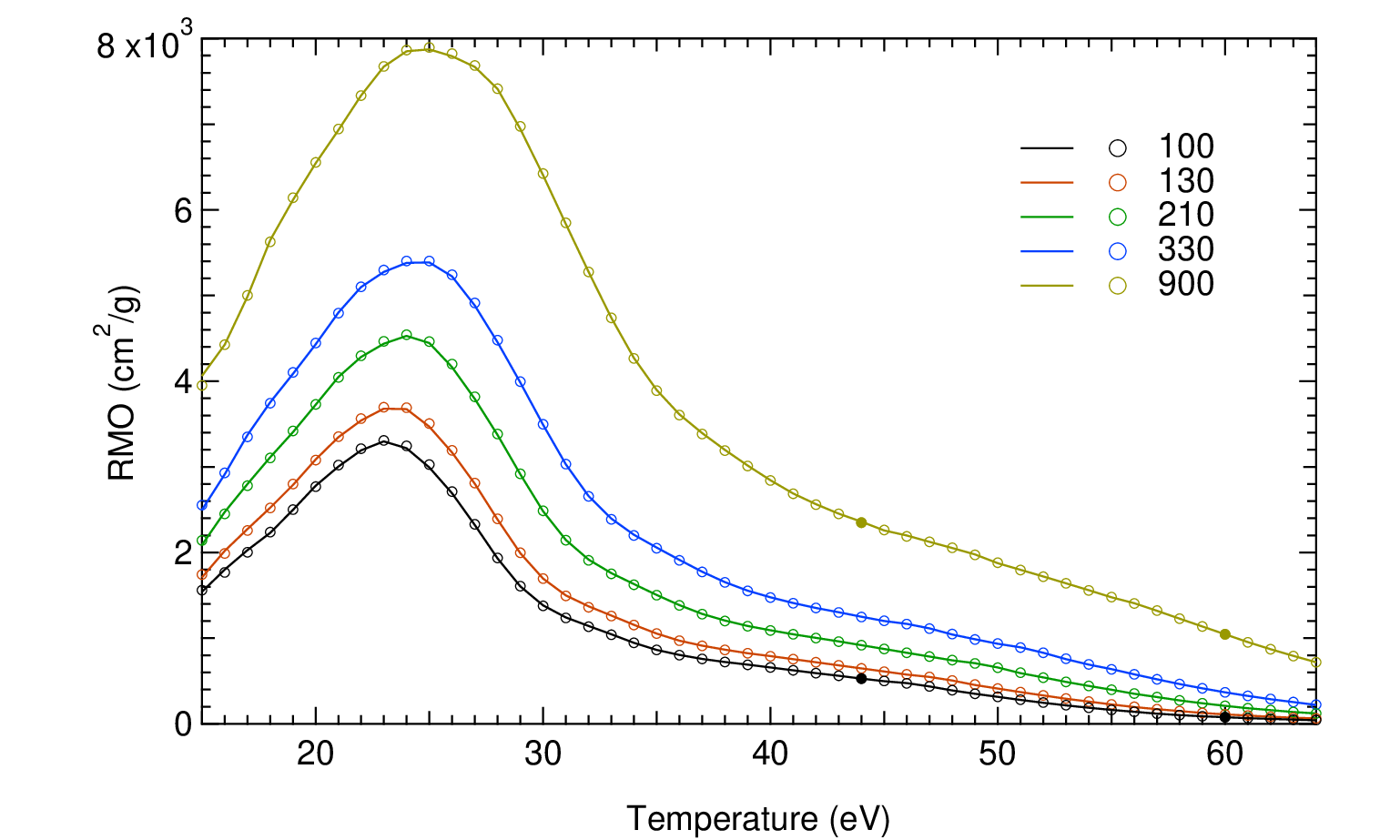}
\caption{Rosseland mean opacity of nickel, as a function of temperature for five mass densities (in $\mu$g/cm$^3$). Lines represent calculated values; open circles indicate predicted values; filled circles correspond to the cases shown in Figs. \ref{OPA_Ni_low_2} and \ref{OPA_Ni_high}.}\label{Pred_Nickel_Rosseland}
\end{figure}

\begin{figure}[htpb]
\centering
\includegraphics[scale=.5]{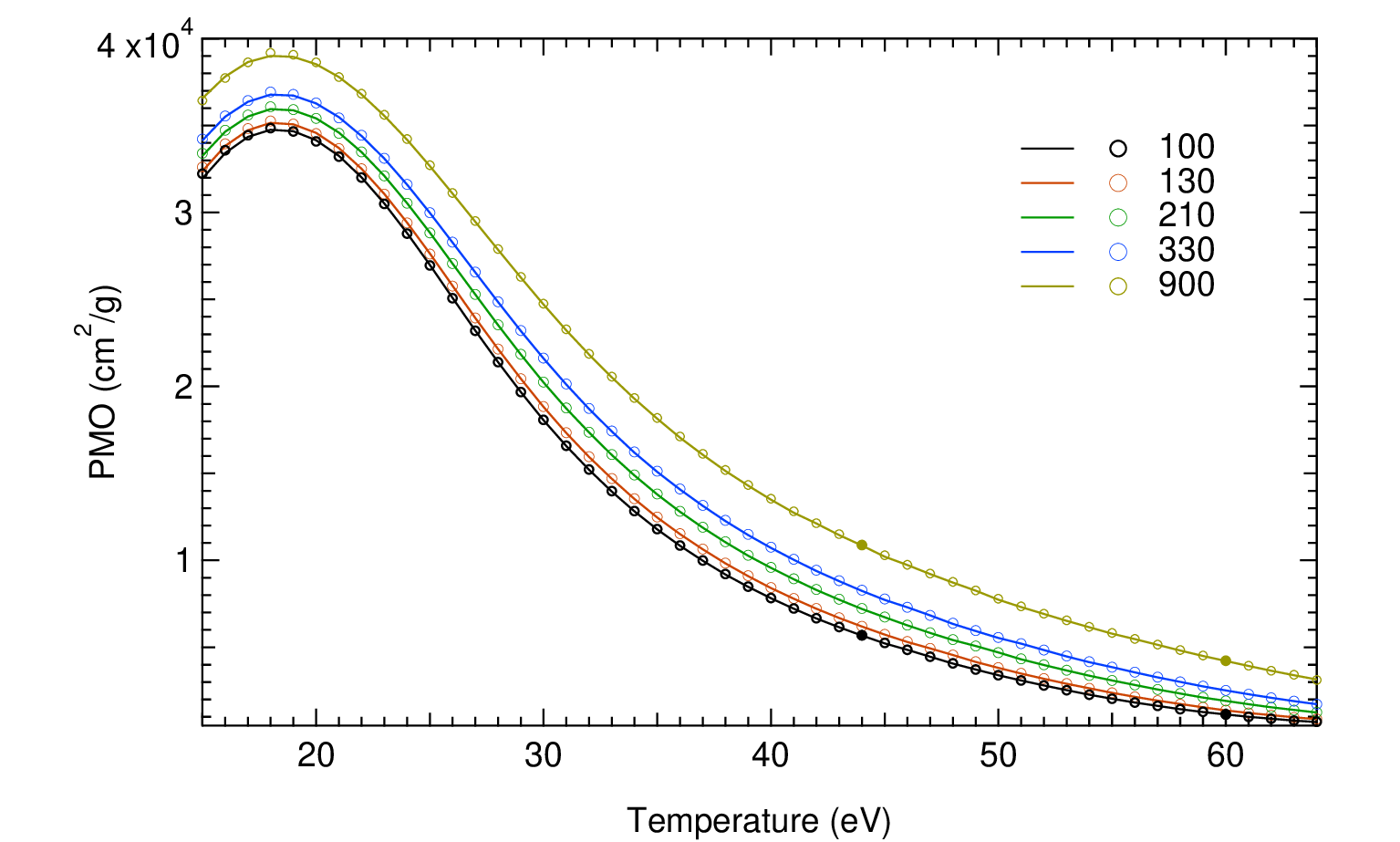}
\caption{Planck mean opacity of nickel as a function of temperature for five mass densities (in $\mu$g/cm$^3$). Lines represent calculated values; open circles indicate predicted values; filled circles correspond to the cases shown in Figs. \ref{OPA_Ni_low_2} and \ref{OPA_Ni_high}.}\label{Pred_Nickel_Planck}
\end{figure}

Figure \ref{Pred_Fer_Rosseland} illustrates the temperature dependence of the Rosseland mean opacity of iron for five representative densities. The predicted values (denoted by the index $p$ in the following) closely match the mean opacities calculated in Sec. \ref{mean opacities} (denoted by the index $c$ in the following), over the whole temperature-density dataset. Model performance is evaluated using the relative difference $|\kappa_R^c-\kappa_R^p|/\kappa_R^c$, quantified by the mean absolute relative error (MARE), the maximum absolute relative error, and the standard deviation (SD) (see Table \ref{tab:statistics} and Appendix \ref{Loss}). These metrics, together with the low loss value of $3.47\times10^{-6}$, demonstrate highly effective training.
Figure \ref{Pred_Fer_Planck} shows the variation of the Planck mean opacity over the same range of plasma parameters. The predicted values closely reproduce the calculated opacity, as quantified by the MARE, and the standard deviation (see Table \ref{tab:statistics}). Again, these metrics, together with a loss value of $2.58\times10^{-6}$, demonstrates the effectiveness of the learning model. Notably, the standard deviation is even smaller than that obtained for the Rosseland mean opacity.
\\
\indent Let us now concentrate on nickel. Figure \ref{Pred_Nickel_Rosseland} represents the variation of $\kappa_R$ with temperature for five densities ranging from 100 to 900 $\mu$g/cm$^3$. Unlike iron, the opacity increases with temperature, reaches a maximum in the interval 20$-$25 eV, and then decreases rapidly. As in the iron case, the CNN-based predictions are in excellent agreement with the values computed using the opacity code described in Section \ref{sec:Opacity}. The MARE and loss values are higher than those obtained for iron, although they remain satisfactory. This is attributable to the greater number of peaks in the low-energy region (see Fig. \ref{OPA_Ni_low_2}). We notice the same level of agreement between the calculated and predicted mean opacities for nickel as for iron, as shown in Figs. \ref{Pred_Nickel_Rosseland} and \ref{Pred_Nickel_Planck}. Overall, these results demonstrate the high accuracy and efficiency of the predictions.
Figure \ref{Pred_Nickel_Planck} represents the variation of $\kappa_P$ with temperature, for the same densities as in Fig. \ref{Pred_Nickel_Rosseland}. The agreement between calculated and predicted values is very satisfactory. The statistical indicators (see Table \ref{tab:statistics}) show that the MARE and standard deviation are higher than for iron, while the loss remains very close. 

\begin{table}[ht!]
    \centering
    \caption{Mean and maximum absolute relative errors, as well as the standard deviation, together with the corresponding loss values (see Appendix \ref{Loss}) are reported. The first entry is computed over the five selected densities and all temperatures, while the second entry (in parentheses) is computed over all densities and temperatures for both iron and nickel.}
    \label{tab:statistics}
    \begin{tabular}{lcccc}
        \toprule
        & \multicolumn{2}{c}{Iron} & \multicolumn{2}{c}{Nickel} \\
        \cmidrule(lr){2-3} \cmidrule(lr){4-5}
        Metric & Rosseland & Planck & Rosseland & Planck \\
        \midrule
        MARE     & 2.78 (2.56)$\times10^{-3}$ & 1.54 (1.70)$\times10^{-3}$ 
                 & 3.48 (3.12)$\times10^{-3}$ & 2.19 (2.78)$\times10^{-3}$ \\
        Maximum  & 4.49 (4.49)$\times10^{-2}$ & 1.68 (5.51)$\times10^{-2}$ 
                 & 2.55 (8.98)$\times10^{-2}$ & 1.75 (2.75)$\times10^{-2}$ \\
        SD       & 3.98 (3.32)$\times10^{-3}$ & 1.94 (2.62)$\times10^{-3}$ 
                 & 3.62 (4.56)$\times10^{-3}$ & 2.17 (2.39)$\times10^{-3}$ \\
        Loss     & 3.47$\times10^{-6}$ & 2.58$\times10^{-6}$ 
                 & 6.21$\times10^{-6}$ & 2.62$\times10^{-6}$ \\
        \bottomrule
    \end{tabular}
\end{table}

\clearpage 
\section{Conclusion}
We predicted opacity spectra for iron and nickel over wide ranges of temperature and density. In this aim, we used multilayer perceptron and convolutional neural network models, applied either independently or in a hybrid configuration to improve predictive performance. For iron, the predicted opacity shows excellent agreement with direct calculations over the entire spectral range, including the K-, L-, and M-shell. For nickel, the predictions also agree well with direct calculations in the K-shell region across all temperatures and densities. In the L- and M-shell regions, the agreement remains satisfactory except at high temperatures, where performance degrades due to a substantial increase in the number of contributing transitions.

Predictions of Rosseland and Planck mean opacities obtained with a one-dimensional CNN model show very good agreement with direct calculations from the {\sc Iliade} code across all temperatures and densities, for both iron and nickel. 

A key challenge in deep learning is that increasing the number of epochs does not necessarily improve predictive accuracy. This necessitates a compromise between computational cost and training performance. Accordingly, prediction accuracy is assessed using multiple metrics, including the mean squared error (MSE), mean and maximum absolute relative errors, standard deviation, and loss values.

We are also investigating alternative training architectures originally developed for classification and detection tasks, such as recurrent neural networks (RNNs). The main challenge lies in adapting these models to handle energy-dependent functions whose shapes vary significantly with plasma conditions.




\section*{Appendix A: Multilayer Perceptron architecture}\label{MLP}
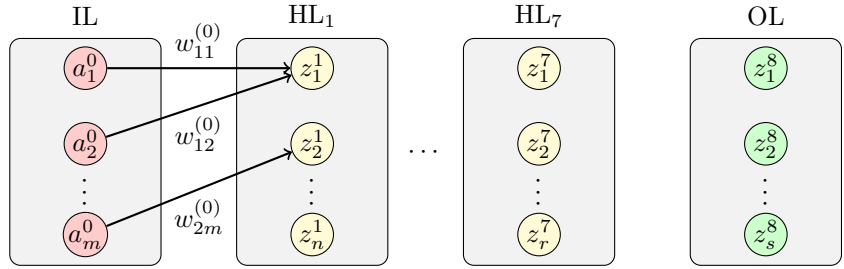
\begin{figure}[htpb]
\centering
\begin{tikzpicture}

\tikzstyle{layer} = [rectangle, rounded corners, draw, fill=gray!10,
text centered, minimum height=3cm, minimum width=2cm]

\tikzstyle{neuron} = [circle, draw, minimum size=16pt, inner sep=0pt]
\tikzstyle{input_neuron} = [neuron, fill=red!20]
\tikzstyle{hidden_neuron} = [neuron, fill=yellow!20]
\tikzstyle{output_neuron} = [neuron, fill=green!20]

\node (IL) [layer] at (5,0)  {};
\node (HL1) [layer] at (8,0)  {};
\node (HL7) [layer] at (11,0) {};
\node (OL) [layer] at (14,0) {};

\node at (5,1.8) {IL};
\node at (8,1.8) {HL$_1$};
\node at (11,1.8) {HL$_7$};
\node at (14,1.8) {OL};

\node at (9.5,0) {$\cdots$};

\node (I1) [input_neuron] at (IL.north) [yshift=-0.4cm] {$a^0_1$};
\node (I2) [input_neuron] at (IL.north) [yshift=-1.4cm] {$a^0_2$};
\node (I3) [input_neuron] at (IL.north) [yshift=-2.6cm] {$a^0_m$};
\node[rotate=90] at (5,-.5) {$\cdots$};

\node (H11) [hidden_neuron] at (HL1.north) [yshift=-0.4cm] {$z^1_1$};
\node (H12) [hidden_neuron] at (HL1.north) [yshift=-1.4cm] {$z^1_2$};
\node (H13) [hidden_neuron] at (HL1.north) [yshift=-2.6cm] {$z^1_n$};
\node[rotate=90] at (8,-.5) {$\cdots$};

\node (H71) [hidden_neuron] at (HL7.north) [yshift=-0.4cm] {$z^7_1$};
\node (H72) [hidden_neuron] at (HL7.north) [yshift=-1.4cm] {$z^7_2$};
\node (H73) [hidden_neuron] at (HL7.north) [yshift=-2.6cm] {$z^7_r$};
\node[rotate=90] at (11,-.5) {$\cdots$};

\node (O1) [output_neuron] at (OL.north) [yshift=-0.4cm] {$z^8_1$};
\node (O2) [output_neuron] at (OL.north) [yshift=-1.4cm] {$z^8_2$};
\node (O3) [output_neuron] at (OL.north) [yshift=-2.6cm] {$z^8_s$};
\node[rotate=90] at (14,-.5) {$\cdots$};

\draw[->, thick] (I1) -- (H11) node[midway, above] {$w^{(0)}_{11}$};
\draw[->, thick] (I2) -- (H11) node[midway, below] {$w^{(0)}_{12}$};
\draw[->, thick] (I3) -- (H12) node[midway, below] {$w^{(0)}_{2m}$};

\end{tikzpicture}

\caption{Multilayer perceptron architecture.
IL: input layer, HL$_i$: hidden layers, OL: output layer. Nodes represent the pre-activations $z^l$; the corresponding activations are $a^l = f(z^l)$.}
\label{fig:MLP_architecture}
\end{figure}

A multilayer perceptron is a feedforward artificial neural network composed of successive fully connected layers (see Fig. \ref{fig:MLP_architecture}) that learn nonlinear relationships between input and output variables. Thanks to its dense connectivity and nonlinear activation functions, an MLP can approximate complex mappings and capture global dependencies in the data. This architecture is particularly effective for supervised learning tasks involving structured numerical features, offering flexibility, robustness, and strong generalization capabilities when properly trained.


Let $x = (x_1, x_2, \dots, x_m)^\top \in \mathbb{R}^m$ denote the input feature vector. The input layer activations are defined as
$$
a^{0} = x
\Longleftrightarrow
a^{0}_i = x_i, \quad i = 1,\dots,m.
$$
For each layer $l \geq 0$, the transition from layer $l$ to layer $l+1$ is defined by an affine transformation followed by a non-linear activation.
\paragraph{Pre-activation.}
The pre-activation gathers the weighted information transmitted by the previous layer. It corresponds to the linear combination of the activations from the previous layer, weighted by the trainable parameters and shifted by a bias term. It represents the input signal received by each neuron before the application of the non-linear activation function.

Let layer $l$ have $p$ neurons and layer $l+1$ have $q$ neurons. The pre-activation vector $z^{l+1} \in \mathbb{R}^q$ is given by
\[
\begin{pmatrix}
z^{l+1}_1 \\
z^{l+1}_2 \\
\vdots \\
z^{l+1}_{q}
\end{pmatrix}
=
\begin{pmatrix}
w^l_{11} & w^l_{12} & \cdots & w^l_{1p} \\
w^l_{21} & w^l_{22} & \cdots & w^l_{2p} \\
\vdots & \vdots & \ddots & \vdots \\
w^l_{q1} & w^l_{q2} & \cdots & w^l_{qp}
\end{pmatrix}
\begin{pmatrix}
a^l_1 \\
a^l_2 \\
\vdots \\
a^l_{p}
\end{pmatrix}
+
\begin{pmatrix}
b^l_1 \\
b^l_2 \\
\vdots \\
b^l_{q}
\end{pmatrix},
\]
or, equivalently,
\begin{equation*}
z^{l+1} = W^l a^l + b^l,
\end{equation*}
where $a^l$ is the activation vector of layer $l$, $W^l$ is the weight matrix connecting layer $l$ to layer $l+1$, and $b^l$ is the bias vector applied to the neurons of layer $l+1$. This operation corresponds to the fully connected layers illustrated in Fig.~\ref{fig:MLP_architecture}.
\paragraph{Activation functions:}
The activation functions convert the pre-activation signals into nonlinear responses, allowing the network to capture complex relationships that cannot be described by linear transformations alone. The activation vector $a^{l+1} \in \mathbb{R}^q$ is obtained by applying the activation function element-wise to the pre-activation vector $z^{l+1}$, thereby introducing the nonlinearity required for the multilayer perceptron to approximate complex input--output mappings.

The activation vector $a^{l+1} \in \mathbb{R}^q$ can the be expressed as:

\begin{equation*}
a^{l+1} =
\begin{pmatrix}
f(z^{l+1}_1) \\
f(z^{l+1}_2) \\
\vdots \\
f(z^{l+1}_{q})
\end{pmatrix}
= f\!\left(z^{l+1}\right).
\end{equation*}
For the hidden layers ($l = 1, \dots, 7$), the activation function $f$ is chosen as the Rectified Linear unit (ReLU):
\begin{equation*}
\mathrm{ReLU}(z) = \max(0, z),
\end{equation*}
where negative inputs are mapped to zero while positive inputs left unchanged. This activation function introduces non-linearity into the network and helps mitigate vanishing-gradient effects during training. For the output layer ($L = 8$), the pre-activations $z^8$ are transformed using a positive activation function in order to guarantee strictly positive predictions, as required by the logarithmic loss function.
\section*{Appendix B: Hybrid CNN--MLP--FiLM Architecture}\label{CNN-MLP-FILM}
\begin{figure}[ht]
\centering
\begin{tikzpicture}[
    font=\small,
    >=latex,
    block/.style={draw, thick, rounded corners, align=center, minimum height=1.3cm, minimum width=2.cm},
    input/.style={block, fill=blue!10},
    conv/.style={block, fill=orange!20},
    res/.style={block, fill=orange!20},
    film/.style={block, fill=purple!20},
    mlp/.style={block, fill=red!10},
    output/.style={block, fill=green!10},
    arrow/.style={->, thick},
]

\node[input] (x) at (0,0)
{Input\\
$E,\;T,\;\rho$\\
 $3 \times L$};

\node[conv] (c1) at (3.,0)
{Conv1D + BN \\+ ReLU\\
 $3 \rightarrow C$};

\node[res] (res) at (6.4,0)
{Residual CNN \\blocks\\
 $C \rightarrow C$};

\node[film] (film) at (10.5,0)
{FiLM modulation\\
$\gamma(T,\rho)\odot x + \beta(T,\rho)$\\
$C \times L$};

\node[conv] (c2) at (10.5,-2.5)
{Conv1D output\\
 $C \rightarrow 1$};

\node[output][align=center, minimum width=3cm, minimum height=1.3cm] (y) at (7.,-2.5)
{Output: $\kappa(E)$\\
$1 \times L$};

\draw[arrow] (x) -- (c1);
\draw[arrow] (c1) -- (res);
\draw[arrow] (res) -- (film);
\draw[arrow] (film) -- (c2);
\draw[arrow] (c2) -- (y);

\node[input] (tr) at (5.,2.5)
{Parameters\\
$T,\rho$};

\node[mlp] (m1) at (7.5,2.5)
{MLP\\
$2 \rightarrow H$};

\node[film] (gb) at (10.5,2.5)
{FiLM parameters\\
$\gamma,\beta \in \mathbb{R}^{C}$};

\draw[arrow] (tr) -- (m1);
\draw[arrow] (m1) -- (gb);
\draw[arrow] (gb) -- (film);

\end{tikzpicture}
\caption{Architecture of the hybrid CNN--MLP--FiLM model.}\label{fig:mlp_cnn_film}
\end{figure}
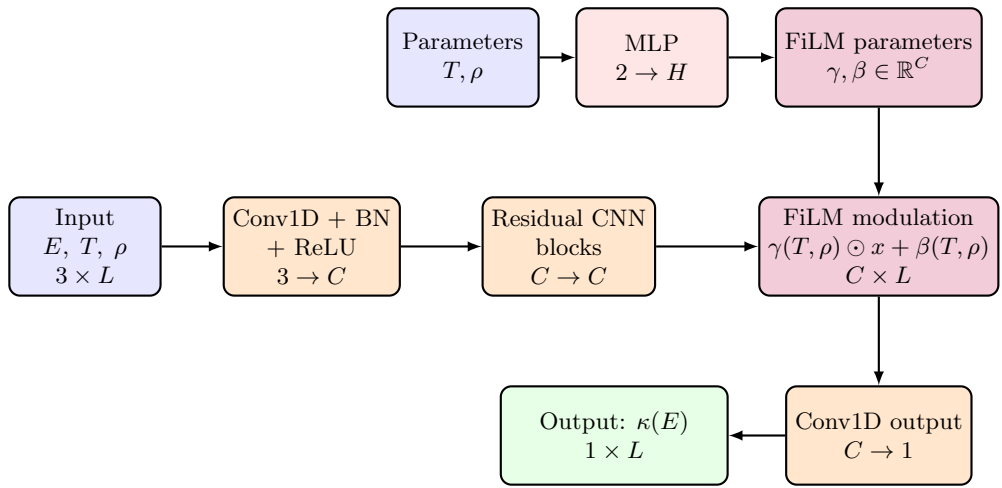

While MLP efficiently captures global nonlinear relationships between the input parameters and the predicted quantities, it does not explicitly exploit the local spectral structures present in the opacity spectra. The hybrid CNN--MLP--FiLM architecture combines the strengths of the convolutional neural network --CNN-- and MLP: the CNN extracts localized and multi-scale features along the energy axis, while the FiLM conditioning mechanism dynamically modulates these features using the plasma parameters learned by the MLP branch. This combination improves the representation of complex spectral patterns, enhances robustness to noise and line blending, and ultimately provides in many cases more accurate predictions than the MLP architecture alone.
Figure \ref{fig:mlp_cnn_film} illustrates the hybrid architecture used to improve opacity predictions. The rectangular blocks denote tensor-valued feature maps rather than individual neurons. The multilayer perceptron processes the plasma parameters $(T,\rho)$ and conditions the convolutional network via FiLM modulation. Tensor shapes are defined as follows: $L$ is the length of the energy grid; the input has three channels, $[E, T, \rho]$, with $T$ and $\rho$ broadcast along the grid to match the sequence length; $C$ denotes the number of convolutional feature maps (64 in the present implementation); and $H$ is the MLP hidden dimension (128 in the present implementation).

We now describe the roles of the individual blocks in Fig.~\ref{fig:mlp_cnn_film}.

\begin{enumerate}[label=(\alph*)]
\item For each plasma state defined by $T$ and $\rho$ (first block) the neural network predicts the opacity profile $\kappa(E)$ sampled on a fixed energy grid (final block, Output). 
\item Each convolutional block consists of a one-dimensional convolution (Conv1D), followed by batch normalization (BN) and a Rectified Linear Unit (ReLU) activation. The Conv1D layer extracts local features along the energy dimension, BN stabilizes and accelerates training by normalizing intermediate activations, and ReLU introduces non-linearity into the model by suppressing negative values. The mapping $3 \rightarrow C$ corresponds to the first Conv1D layer, which projects the three input channels ($E$, $T$, $\rho$) onto $C$ feature maps.
\item The transformation $C \rightarrow C$ refers to residual convolutional blocks that maintain the same number of feature maps while refining and combining learned representations without changing the tensor dimensionality.
\item FiLM modulation: a tensor of size $C \times L$ represents a set of $C$ learned feature maps defined along the energy grid of length $L$, where each feature map encodes a different aspect of the spectral structure.
\item Conv1D output: the operation $C \rightarrow 1$ denotes the final convolutional layer, which compresses all feature maps into a single output channel representing the predicted quantity.
\item Output: the tensor $1 \times L$ corresponds to the predicted opacity profile $\kappa(E)$ evaluated on the energy grid $E$.
\end{enumerate}

After presenting an overview of the various blocks and their interconnections, we turn to a discussion of how the ensemble works in practice.

\begin{enumerate}
\item Input representation\\
Each training sample is defined on a discretized energy grid $E$, together with a set of plasma parameters $(T,\rho)$ and an associated
target function $\kappa(E)$.

\item Role of the CNN\\
The convolutional neural network is designed to extract features localized in the energy domain and therefore operates exclusively along the energy dimension. Through successive one-dimensional convolutional layers (Conv1D+Batch Normalization+ReLU activations), including residual connections, the network learns local spectral structures such as sharp variations, characteristic widths, spectral peaks, and smooth background components. Peak localization emerges naturally from the convolutional architecture, whose filters act as learned local detectors responding to variations in the spectrum, while weight sharing provides translation invariance along the energy axis. Furthermore, the hierarchical multi-layer structure enables multi-scale feature extraction: early layers capture narrow spectral lines, whereas deeper layers integrate broader spectral context, improving robustness to noise and line blending.

Residual CNN blocks are used to enhance training stability and enable the network to learn incremental refinements of previously extracted features. Following the convolutional stages, the CNN outputs a latent representation
\[
\mathbf{H} \in \mathbb{R}^{C \times L}.
\]
This tensor can be interpreted as a set of $C$ learned feature maps defined over the energy grid of length $L$, each encoding different local spectral patterns such as peaks, edges, or smooth background variations. 
At this stage, the representation $\mathbf{H}$ depends only on the input spectrum and is therefore independent of the plasma parameters $(T,\rho)$. These physical parameters are introduced later in the network and used to condition or modulate $\mathbf{H}$, ensuring that the final prediction accounts for the thermodynamic state of the plasma.

\item Role of the MLP\\
In parallel, a multilayer perceptron processes the plasma parameters. Its purpose is not to directly predict the opacity profile $\kappa(E)$, but rather to encode the thermodynamic state of the plasma into a compact feature vector. In other words, the MLP learns a nonlinear representation of the input parameters that summarizes their influence in a reduced form. Formally, it implements the mapping
$$(T,\rho) \;\longrightarrow\; \mathbf{Z},$$
where $\mathbf{Z}$ is a latent conditioning vector from which the FiLM modulation parameters are subsequently computed.

\item FiLM conditioning\\
The coupling between the CNN and the MLP is achieved using feature-wise linear modulation (FiLM). From the MLP output, two vectors are generated, $\boldsymbol{\gamma}(T,\rho)$ and $\boldsymbol{\beta}(T,\rho)$ (see Fig.~\ref{fig:mlp_cnn_film}), each of dimension $C$, corresponding to the number of CNN feature channels. For each channel $c$ ($c = 1,\dots,C$), FiLM modulates the CNN features according to
\[
\tilde{\mathbf{H}}_c(E)
=
\gamma_c(T,\rho)\,\mathbf{H}_c(E)
+
\beta_c(T,\rho).
\]
In this relation, $\mathbf{H}_c$ is the original feature values produced by the CNN for channel $c$, while 
$\tilde{\mathbf{H}}_c(E)$ is the modulated feature map after FiLM conditioning
(obtained by scaling and shifting $\mathbf{H}_c(E)$) and then depending on both the CNN features and the thermodynamic parameters. 
\\
This operation preserves the energy-dependent structure learned by the CNN while allowing the plasma parameters to rescale and shift each feature channel. Thus, the CNN learns what features exist in the energy domain, whereas the MLP--FiLM pathway learns how these features depend on thermodynamic conditions.

\item Final prediction\\
After FiLM modulation, a final one-dimensional convolution maps the conditioned features to a single output channel corresponding to $\kappa(E)$. The entire architecture is trained end-to-end by minimizing a mean squared error loss.
\end{enumerate}

In conclusion, the model decomposes the learning task as follows: the CNN captures universal energy-domain structures, the MLP encodes thermodynamic state information, FiLM acts as a physically meaningful interface allowing $(T,\rho)$ to continuously deform the energy-dependent response. This design is particularly well suited for problems in which the functional form in $E$ is shared across conditions, but its amplitude, shape, or scale varies smoothly with $(T,\rho)$.

\section*{Appendix C: Loss and numerical errors}\label{Loss}
The neural network is trained by minimizing the mean squared error (MSE) in logarithmic space over both the thermodynamic states ($T$, $\rho$) and the energy grid:

\begin{equation*}
\mathrm{Loss}
=
\frac{1}{N\times L}
\sum_{i=1}^{N}
\sum_{k=1}^{L}
\left[
\ln \kappa^{\mathrm{p}}_{i,k}
-
\ln \kappa^{\mathrm{c}}_{i,k}
\right]^2 ,
\end{equation*}
where $i$ indexes the thermodynamic states $(T,\rho)_i$ and $k$ the energy grid points $E_k$. Here, $\kappa^{\mathrm{p}}$ and $\kappa^{\mathrm{c}}$ denote the predicted and reference values, respectively. The loss therefore measures the quadratic deviation of the natural logarithm of the opacity. The corresponding root-mean-square (RMS) error in logarithmic space is

\begin{equation*}
\sigma_{\log}^{\mathrm{RMS}} = \sqrt{\mathrm{Loss}}.
\end{equation*}
The MSE can be decomposed into variance and bias contributions,

\begin{equation*}
\mathrm{Loss}
=
\mathrm{Var}(\epsilon_{\log})
+
\overline{\epsilon}_{\log}^{\,2},
\end{equation*}
where

\begin{equation*}
\overline{\epsilon}_{\log}
=
\frac{1}{N\times L}
\sum_{i=1}^N\sum_{k=1}^L
\left(
\ln\kappa^{\mathrm{p}}_{i,k}
-
\ln\kappa^{\mathrm{c}}_{i,k}
\right)
\end{equation*}
is the mean logarithmic error.

For small relative deviations 
$\varepsilon = (\kappa^{\mathrm{p}}-\kappa^{\mathrm{c}})/\kappa^{\mathrm{c}}$,
one has
\[
\ln(1+\varepsilon) \simeq \varepsilon,
\]
so that the RMS error in logarithmic space provides, to first order, an estimate of the RMS relative error in linear space. This approximation holds provided that the relative error remains sufficiently small over most of the dataset.

\end{document}